\documentclass[11pt,a4paper]{JHEP3}
\usepackage{graphicx}
\usepackage{epstopdf}
\usepackage{cite}
\usepackage{axodraw}
\usepackage{multicol}
\newcommand{\LMSSM}{ {\slash\hspace*{-0.24cm}}L-MSSM~}
\newcommand{\xib}{\bar{\xi}}
\newcommand{\Jb}{\bar{J}}
\newcommand{\Lag}{\mathcal{L}}
\newcommand{\mama}[1]{\mathcal{M}_{#1}}

\newcommand{\SPot}{\mathcal{W}}

\newcommand{\half}{\frac{1}{2}}

\newcommand{\tar}[1]{\bar{\tilde{#1}}}
\newcommand{\isqt}{\frac{1}{\sqrt{2}}}
\newcommand{\ep}[1]{ \varepsilon^{\phantom{*}}_{#1} }

\newcommand{\YU}[1]{\left( Y_U   \right)^{\phantom{*}}_{#1}}
\newcommand{\YUs}[1]{ \left( Y_U \right)^{*}_{#1} }

\newcommand{\lam}[1]{ \lambda^{\phantom{*}}_{#1} }
\newcommand{\lams}[1]{ \lambda^*_{#1} }
\newcommand{\lamp}[1]{ {\lambda'}^{\phantom{*}}_{\! \!  #1} }
\newcommand{\lamps}[1]{ {\lambda'}^*_{\! \! #1} }

\newcommand{\mux}[1]{\mu_{#1}}
\newcommand{\muxs}[1]{\mu^*_{#1}}

\newcommand{\sigb}[2]{  \bar{\sigma}_{#1}^{#2}  }
\newcommand{\Q}[3]{ Q_{#1}^{#2 #3} }

\newcommand{\sQ}[2]{ \tilde{Q}_{#1}^{#2} }
\newcommand{\uL}[3]{ u^{#2 #3}_{L #1} }
\newcommand{\dL}[3]{ d^{#2 #3}_{L #1} }
\newcommand{\uLb}[3]{ \bar{u}^{#2}_{L #1 #3} }
\newcommand{\dLb}[3]{ \bar{d}^{#2 }_{L #1 #3} }
\newcommand{\suL}[2]{ \tilde{u}^{#2}_{L #1} }
\newcommand{\suLs}[2]{ \tilde{u}^{* #2 }_{L #1} }
\newcommand{\sdL}[2]{ \tilde{d}^{#2}_{L #1} }
\newcommand{\sdLs}[2]{ \tilde{d}^{* #2 }_{L #1} }
\newcommand{\sUc}[2]{ \tilde{U}_{#1}^{c #2} }
\newcommand{\sDc}[2]{ \tilde{D}_{#1}^{c #2} }
\newcommand{\uR}[3]{ u^{#2 #3}_{R #1} }
\newcommand{\uRb}[3]{ \bar{u}^{#2 }_{R #1 #3} }
\newcommand{\dR}[3]{ d^{#2 #3}_{R #1} }
\newcommand{\dRb}[3]{ \bar{d}^{\, #2}_{R #1 #3} }
\newcommand{\suR}[2]{ \tilde{u}^{#2}_{R #1} }
\newcommand{\suRs}[2]{ \tilde{u}^{* #2 }_{R #1} }
\newcommand{\sdR}[2]{ \tilde{d}^{#2}_{R #1} }
\newcommand{\sdRs}[2]{ \tilde{d}^{* #2 }_{R #1} }
\newcommand{\Lx}[2]{ \mathcal{L}_{#1}^{#2} }

\newcommand{\eL}[2]{ e^{#2}_{L #1} }
\newcommand{\eLb}[2]{ \bar{e}^{\phantom{*}}_{L #1 #2} }
\newcommand{\nuL}[2]{ \nu^{#2}_{L #1} }

\newcommand{\seL}[1]{ \tilde{e}^{\phantom{*}}_{L #1}}
\newcommand{\seLs}[1]{ \tilde{e}^{*}_{L #1} }

\newcommand{\snuL}[1]{ \tilde{\nu}^{\phantom{*}}_{L #1} }
\newcommand{\snuLs}[1]{ \tilde{\nu}^{*}_{L #1} }

\newcommand{\snuLvev}[1]{ v_{#1} }
\newcommand{\snuLsvev}[1]{ v_{#1} }

\newcommand{\sEc}[2]{ \tilde{E}_{#1}^{c \, #2} }
\newcommand{\eR}[2]{ e^{#2}_{R #1} }
\newcommand{\eRb}[2]{ \bar{e}^{\phantom{*}}_{R #1 #2} }

\newcommand{\seR}[1]{ \tilde{e}^{\phantom{*}}_{R #1} }
\newcommand{\seRs}[1]{ \tilde{e}^{*}_{R #1} }

\newcommand{\HB}[2]{ {H_2}_{#1}^{#2} }

\newcommand{\hinobz}[1]{ \tilde{h}_2^{0 #1} }
\newcommand{\hinobp}[1]{ \tilde{h}_2^{+ #1} }

\newcommand{\hinobpb}[1]{ \tar{h}_{2 #1}^{+} }
\newcommand{\sham}{ h_1^- }
\newcommand{\shams}{ h_1^{-*} }
\newcommand{\shbz}{ h_2^0 }
\newcommand{\shbzs}{ h_2^{0*} }

\newcommand{\shbzvev}{ v_u }
\newcommand{\shbzsvev}{ v_u }

\newcommand{\shbp}{ h_2^+ }
\newcommand{\shbps}{ h_2^{+*} }

\newcommand{\W}[2]{W_{#1}^{#2}}
\newcommand{\Z}[1]{Z_{#1}}

\newcommand{\Bino}[2]{\widetilde{B}_{#1}^{#2}}
\newcommand{\Wino}[2]{\widetilde{W}_{#1}^{#2}}
\newcommand{\Gino}[2]{\widetilde{G}_{#1}^{#2}}

\newcommand{\Winob}[2]{\bar{\widetilde{W}}_{#1}^{#2}}

\newcommand{\M}[2]{ {\cal{M}}_{\tiny{\textup{#1}} #2  }}
\newcommand{\Ms}[2]{ {\cal{M}}^{*}_{\tiny{\textup{#1}} #2  }}
\newcommand{\Msq}[2]{ {\cal{M}}^2_{\tiny{\textup{#1}}#2  }}

\newcommand{\Mh}[2]{ {\hat{\cal{M}}}_{\tiny{\textup{#1}} #2 }}
\newcommand{\Msh}[2]{ {\hat{\cal{M}}}^{*}_{\tiny{\textup{#1}} #2 }}
\newcommand{\Msqh}[2]{ {\hat{\cal{M}}}^2_{\tiny{\textup{#1}} #2 }}

\newcommand{\ZH}[1]{ Z_{H#1} }
\newcommand{\ZR}[1]{ Z_{R#1} }
\newcommand{\ZA}[1]{ Z_{A#1} }
\newcommand{\Zp}[1]{ Z_{+#1} }
\newcommand{\Zm}[1]{ Z_{-#1} }
\newcommand{\ZN}[1]{ Z_{N#1} }

\newcommand{\Zsu}[1]{ Z_{\tilde{u}#1} }
\newcommand{\Zsd}[1]{ Z_{\tilde{d}#1} }

\newcommand{\ZdR}[1]{ Z_{d_R#1} }
\newcommand{\ZuL}[1]{ Z_{u_L#1} }

\newcommand{\ZHs}[1]{ Z^*_{H#1} }
\newcommand{\ZRs}[1]{ Z^*_{R#1} }
\newcommand{\ZAs}[1]{ Z^*_{A#1} }
\newcommand{\Zps}[1]{ Z^*_{+#1} }
\newcommand{\Zms}[1]{ Z^*_{-#1} }
\newcommand{\ZNs}[1]{ Z^*_{N#1} }

\newcommand{\Zsus}[1]{ Z^*_{\tilde{u}#1} }
\newcommand{\Zsds}[1]{ Z^*_{\tilde{d}#1} }
\newcommand{\ZdLs}[1]{ Z^*_{d_L#1} }

\newcommand{\ZuLs}[1]{ Z^*_{u_L#1} }
\newcommand{\ZuRs}[1]{ Z^*_{u_R#1} }
\newcommand{\ZHd}[1]{ Z^\dagger_{H#1} }
\newcommand{\ZRd}[1]{ Z^\dagger_{R#1} }
\newcommand{\ZAd}[1]{ Z^\dagger_{A#1} }
\newcommand{\Zpd}[1]{ Z^\dagger_{+#1} }
\newcommand{\Zmd}[1]{ Z^\dagger_{-#1} }
\newcommand{\ZNd}[1]{ Z^\dagger_{N#1} }

\newcommand{\Zsud}[1]{ Z^\dagger_{\tilde{u}#1} }

\newcommand{\ZRT}[1]{ Z^T_{R#1} }
\newcommand{\ZAT}[1]{ Z^T_{A#1} }
\newcommand{\ZpT}[1]{ Z^T_{+#1} }
\newcommand{\ZmT}[1]{ Z^T_{-#1} }
\newcommand{\ZNT}[1]{ Z^T_{N#1} }

\newcommand{\ZsdT}[1]{ Z^T_{\tilde{d}#1} }

\newcommand{\Hp}[1]{ H^+_{ #1 }}
\newcommand{\Hz}[1]{ H^0_{ #1 } }
\newcommand{\Az}[1]{ A^0_{ #1 } }
\newcommand{\Kp}[2]{ \kappa^{+#2}_{#1} }
\newcommand{\Km}[2]{ \kappa^{-#2}_{#1} }
\newcommand{\Kz}[2]{ \kappa^{0#2}_{#1} }

\newcommand{\su}[2]{ \tilde{u}_{#1}^{#2} }
\newcommand{\sd}[2]{ \tilde{d}_{#1}^{#2} }

\newcommand{\Hps}[1]{ H^{+*}_{ #1 }}

\newcommand{\Kpb}[2]{ \bar{\kappa}^{+#2}_{#1} }
\newcommand{\Kmb}[2]{ \bar{\kappa}^{-#2}_{#1} }
\newcommand{\Kzb}[2]{ \bar{\kappa}^{0#2}_{#1} }

\newcommand{\sus}[2]{ \tilde{u}_{#1}^{*#2} }
\newcommand{\sds}[2]{ \tilde{d}_{#1}^{*#2} }

\newcommand{\bea}{\begin{eqnarray}}
\newcommand{\eea}{\end{eqnarray}}

\newcommand{\iin}[8]
{ \begin{tabular}{ll}
  \begin{picture}(150,70)(0,-10)
    \DashLine(60,0)(10,0){5} \Text(0,0)[c]{$#3$} \ArrowLine(110,0)(60,0) \Text(120,0)[c]{$#2$}
    \ArrowLine(60,50)(60,0) \Text(65,45)[l]{$#1$} \Vertex(60,0){2}
  \end{picture} & \raisebox{5\unitlength}
  { \begin{minipage}{5cm} \lefteqn{ #4 } \lefteqn{ #5 } \lefteqn{ #6 } \lefteqn{ #7 } \lefteqn{ #8 } \end{minipage} }
\end{tabular} }

\newcommand{\iisi}[8]
{ \begin{tabular}{ll}
  \begin{picture}(150,70)(0,-10)
    \DashArrowLine(10,0)(60,0){5} \Text(0,0)[c]{$#3$} \ArrowLine(110,0)(60,0) \Text(120,0)[c]{$#2$}
    \ArrowLine(60,50)(60,0) \Text(65,45)[l]{$#1$} \Vertex(60,0){2}
  \end{picture} & \raisebox{15\unitlength}
  { \begin{minipage}{5cm} \lefteqn{ #4 } \lefteqn{ #5 } \lefteqn{ #6 } \lefteqn{ #7 } \lefteqn{ #8 }\end{minipage} }
\end{tabular} }

\newcommand{\iiso}[8]
{ \begin{tabular}{ll}
  \begin{picture}(150,70)(0,-10)
    \DashArrowLine(60,0)(10,0){5} \Text(0,0)[c]{$#3$} \ArrowLine(110,0)(60,0) \Text(120,0)[c]{$#2$}
    \ArrowLine(60,50)(60,0) \Text(65,45)[l]{$#1$} \Vertex(60,0){2}
  \end{picture} & \raisebox{15\unitlength}
  { \begin{minipage}{5cm} \lefteqn{ #4 } \lefteqn{ #5 } \lefteqn{ #6 } \lefteqn{ #7 } \lefteqn{ #8 }\end{minipage} }
\end{tabular} }

\newcommand{\ffv}[5]
{ \begin{tabular}{ll}
  \begin{picture}(150,70)(0,-10)
    \Photon(10,0)(60,0){3}{4} \Text(0,0)[c]{$#3$} \ArrowLine(60,0)(110,0)
    \Text(120,0)[c]{$#1$} \ArrowLine(60,50)(60,0) \Text(65,45)[l]{$#2$} \Vertex(60,0){2}
  \end{picture} & \raisebox{35\unitlength} 
  { \begin{minipage}{5cm} \lefteqn{ #4 } \lefteqn{ #5 } \end{minipage} }
\end{tabular} }

\title{Neutrino masses in the Lepton Number Violating MSSM}

\author{A.  Dedes, S.  Rimmer \\ Institute for Particle Physics
Phenomenology (IPPP), Durham DH1 3LE, UK}

\author{J.  Rosiek \\ Institute of Theoretical Physics, Warsaw
University, Hoza 69, 00-681 Warsaw, Poland}

\preprint{IPPP/06/17 
\\ DCPT/06/34
\\ hep-ph/0603225}

\date{} 

\abstract{We consider the most general supersymmetric model with
minimal particle content and an additional discrete ${\cal Z}_3$
symmetry (instead of R-parity), which allows lepton number violating
terms and results in non-zero Majorana neutrino masses. We investigate
whether the currently measured values for lepton masses and mixing can
be reproduced.
We set up a framework in which Lagrangian parameters can be
initialised without recourse to assumptions concerning trilinear or
bilinear superpotential terms, CP-conservation or intergenerational
mixing and analyse in detail the one loop corrections to the neutrino
masses.  We present scenarios in which the experimental data
are reproduced and show the effect varying lepton number violating
couplings has on the predicted atmospheric and solar mass$^2$
differences.
We find that with bilinear lepton number violating couplings in the
superpotential of the order 1 MeV the atmospheric mass scale can be
reproduced.
Certain trilinear superpotential couplings, usually, of the order of
the electron Yukawa coupling can give rise to either atmospheric or
solar mass scales and bilinear supersymmetry breaking terms of the
order 0.1 $\textup{GeV}^2$ can set the solar mass scale.
Further details of our calculation, Lagrangian, Feynman rules and
relevant generic loop diagrams, are presented in three Appendices.  }

\begin{document}

\section{Introduction}

Once one allows for R-parity~\cite{Fayet} violation in the Minimal
Supersymmetric Standard Model (MSSM) there is an embarrassingly large
class of possible models.  
Building on the seminal paper of Ibanez and Ross~\cite{IR}, it has been shown
that there are three preferred ${\cal Z}_N$ symmetries which can be imposed
when constructing the MSSM with
minimal content of particle fields~\cite{Dreiner:2005}: one is the standard R-parity under
which the Standard Model particles are R-parity even while their
superpartners are R-parity odd, the second is a unique, ${\cal Z}_3$
symmetry which results in the MSSM with lepton number violation;
denoted here as (\LMSSM) and the third is a ${\cal Z}_6$ symmetry refered to as proton hexality.  
The former guarantees a stable lightest
supersymmetric particle and thus missing energy at colliders, the
${\cal Z}_3$ and ${\cal Z}_6$ lead to proton stability\footnote{Baryon number violating
operators of the form $QQQL$ or $\bar{U}\bar{U}\bar{D}\bar{E}$ are not
allowed in \LMSSM.} together with neutrino flavour changing phenomena
and masses.  In this paper we want to investigate in detail neutrino
masses and mixings in the \LMSSM motivated from the current
observations of neutrino flavour metamorphosis and proton
stability~\cite{PDG}.

In the \LMSSM a single neutrino mass arises at tree level due to the
mixing between neutrinos, gauginos and
higgsinos~\cite{Hall,Joshipura,Banks,Nowakowski}.  This tree level
mass is proportional to the bilinear lepton number violating
superpotential parameter, $\mu_i$, squared, which is assumed to be of
order\footnote{This can be naturally accommodated by employing an
R-symmetry\cite{Nilles,Chamseddine:1996rc,Allanach}.} of ${\rm MeV}$, and is suppressed by the
``TeV'' supersymmetry breaking gaugino masses, resulting in a low
energy see-saw mechanism with light neutrino and heavy neutralino
masses.  The other two neutrino masses arise from quantum loop
corrections made up from lepton number violating superpotential or
supersymmetry breaking vertices.  We shall refer to these neutrinos
with the term ``massless neutrinos''.

Calculations for neutrino masses in the \LMSSM have been addressed many
times in the literature.  The tree level set-up of the model was first
given in \cite{Hall}, and details worked out later in \cite{Joshipura,
Banks,Nowakowski}.  Calculations of the one-loop neutrino masses
including only the bilinear superpotential term are given
in~\cite{Hempfling, Kaplan,Valle,ValleD68,Abada}.  Corrections
involving the trilinear superpotential Yukawa couplings $\lambda,
\lambda'$ considered mostly in the mass insertion
approximation~\cite{GHsneutrino,Haber,Grossman,Sacha} and under the
assumption of CP-conservation and flavour diagonal soft SUSY breaking
terms.  Renormalization group induced corrections to neutrino masses
have been studied in~\cite{deCarlos,Nardi,Allanach}.  There is of
course a vast number of articles using these calculations, or
simplified versions of them, in order to describe the solar and
atmospheric neutrino
puzzles~\cite{Borzumati,Drees,Chun,Joshipura2,Choi,Kong,Rakshit:1998kd,Adhikari,Abada2,Rakshit:2004rj}.

In this work, we calculate the complete set of the one-loop
corrections to the massless neutrinos without resorting to
approximations about CP-conservation or bilinear superpotential
operator dominance.  The outline of our paper is as follows : in
section~\ref{sec:2} we show how to define the Lagrangian parameters in
the fermion sector of the theory, by starting out with physical input
parameters like the lepton masses and mixing angles.  In
section~\ref{sec:3} we describe our renormalization procedure and
present analytical results for the one-loop corrections together with
approximate formulas (if necessary) for individual diagrams.  We
compare with the current literature.  In section~\ref{sec:4} we
present numerical results for the size of the input parameters when
these account for the neutrino experimental data, and in
section~\ref{sec:5} draw our conclusions.  Finally, in three
Appendices, we set out our notation, in Appendix~\ref{app:mass}, the
Lagrangian and the mass terms, and present the relevant Feynman rules
in Appendix~\ref{app:feyn}.  Furthermore, in Appendix~\ref{app:weyl}
we present a pedagogical brief introduction to the Weyl spinor
calculation that are employed throughout this paper and present
general one-loop self energy corrections that are employed in this
paper and can be used elsewhere.

The Fortran-code for calculating the neutrino masses used in this
article has been made publicly available\footnote{Please send e-mail
to
\href{mailto:Steven.Rimmer@durham.ac.uk}{Steven.Rimmer@durham.ac.uk}
or \href{mailto:Janusz.Rosiek@fuw.edu.pl}{Janusz.Rosiek@fuw.edu.pl}
for further details regarding the code and guidelines.  The Fortran
source files can be obtained from
\href{http://www.ippp.dur.ac.uk/$\sim$dph3sr/rpv}{http://www.ippp.dur.ac.uk/$\sim$dph3sr/rpv}
or from
\href{http://www.fuw.edu.pl/$\sim$rosiek/rpv/rpv.html}{http://www.fuw.edu.pl/$\sim$rosiek/rpv/rpv.html}}.
It can be used in adding additional constraints when other \- \LMSSM
processes are studied.

\section{Fermion masses and mixings in  \LMSSM }
  \label{sec:2}
  
Fermion masses and couplings of the general \LMSSM are defined by the
superpotential of the model, vacuum expectation values (vevs) of the
neutral scalar fields and the soft supersymmetry breaking gaugino
masses.  The most general superpotential takes the form
\begin{eqnarray}
{\cal W}_{\rm ~ \LMSSM} = \epsilon_{ab}\left( \frac{1}{2}
\lambda_{\alpha\beta k}{\cal L}_{\alpha}^a {\cal L}_{\beta}^b
\bar{E}_k \: + \: \lambda'_{\alpha jk} {\cal L}_{\alpha}^a Q_j^{b,x}
\bar{D}_{k,x} \: + \: (Y_U)_{jk} Q_j^{a,x} H_2^b \bar{U}_{k,x} \: - \:
\mu_\alpha {\cal L}_\alpha^a H_2^b \right ),\nonumber \\
\label{superpot1}
\end{eqnarray}
where $Q_i^{a\,x},\;{\bar D}_i^x,\;{\bar U}_i^x,\; {\cal
L}_i^a,\;{\bar E}_i,\; H_1^a,\;H_2^a$ are the chiral superfield
particle content, $i=1,2,3$ is a generation index, $x=1,2,3$ and
$a=1,2$ are $SU(3) $ and $SU(2)$ gauge indices, respectively.  The
simple form of (\ref{superpot1}) results when combining the chiral
doublet superfields with common hypercharge $Y=-\frac{1}{2}$ into
${\cal L}^a_{\alpha=0,\ldots,3}=(H_1^a,\,L_{i=1,2, 3}^a)$.
$\mu_\alpha$ is the generalized dimensionful $\mu$-parameter, with
$\mu_0$ and $\mu_i, i=1,...3$ the lepton number conserving and
violating parts respectively, and $\lambda_{\alpha \beta
k},\,\lambda'_{\alpha j k}, (Y_U)_{ij}$ are Yukawa matrices with
$\epsilon_{ab}$ being the totally anti-symmetric tensor
$\epsilon_{12}=+1$.

Physical masses of the fermion fields depend on appropriate $\lambda,
\lambda', \mu$ and $Y_U$ couplings multiplied by the vevs of the
neutral scalar fields.  As it has been shown in~\cite{Dreiner:2003hw,DRRS}, by
unitary rotation in the 4-dimensional space of the neutral scalar
components of $\Lx{\alpha}{}$, it is possible to set three of the four
vacuum expectation values of the $\Lx{\alpha}{}$ fields to zero,
leaving two real non-zero vevs in the neutral scalar sector and,
simultaneously, significantly simplifying its structure.  It is
convenient to apply such a transformation not just to scalars, but to
the whole chiral superfield,

\begin{equation} 
\Lx{\alpha}{}=  U_{\alpha \beta} \Lx{\beta}{'}\;,
\label{matU}
\end{equation}
and redefine the Lagrangian parameters to absorb the matrix $\bf U$ in
Eq.~(\ref{matU}) such that it does not appear explicitly in the
Lagrangian,
\bea
\tilde{\lambda}_{\gamma \delta j}&=&\lam{\alpha \beta j}U_{\alpha
\gamma} U_{\beta \delta} \;, \nonumber\\
\tilde{\lambda'}_{\gamma ij} &=& \lamp{\alpha ij} U_{\alpha\gamma} \;,
\\
\tilde{\mu}_\gamma &=& \mux{\alpha} U_{\alpha \gamma}\;.  \nonumber
\eea
The tildes and primes on the fields are then dropped.    

In a standard way, both isospin components of $\Q{}{}{}$ superfield
and of $\bar{U}$, $\bar{D}$ superfields can each be redefined by a
unitary rotation in the flavour space.  As such, it is possible to
diagonalise the Yukawa couplings ${\bf (Y_D)}$, $\bf \YU{}$ (note that
${\bf (Y_D)}\equiv \lambda'_{0ij}$ in the basis with two non-vanishing
scalar vevs) and absorb the rotation matrices in field redefinitions
such that they do not appear explicitly in the Lagrangian, apart from
a specific combination of rotation matrices which appear in the gauge
and Higgs charged currents which is identified as the CKM matrix.  In
this basis, it is clear how to initialise the Lagrangian parameters,
as the diagonal values are then proportional to the measured values
for the up- and down-type quarks.  For more details concerning this
point the reader should consult Appendix~\ref{app:mass}.

In the lepton sector, however, the same approach cannot be adopted for
two reasons.  Firstly, even with a diagonal Yukawa matrix, the charged
lepton masses are given by three eigenvalues of the larger $(5 \times
5)$ mass matrix which includes mixing between the charged fermionic
components of the $\Lx{}{}$, $\bar{E}$ and the charged gauginos and
higgsinos.  Thus, the diagonal entries in the Yukawa matrix would not
correspond exactly to the masses of the physical mass eigenstates
which describe the charged leptons.  Secondly, the $\Lx{}{}$-basis has
already been fixed by the property that three neutral scalar vevs
should be zero, so we are not free to absorb a rotation
matrix\footnote{ This is not entirely true: it is actually possible to
perform a rotation into the vanishing sneutrino vev basis and to
diagonal Yukawa couplings~\cite{OK}; it is possible to use the freedom
in the 3-dimensional lepton space, which we used in Ref.~\cite{DRRS}
to diagonalise the sneutrino masses, in order to make the lepton
Yukawa couplings diagonal.  But then one will have a $10\times 10$
mass matrix for the neutral scalars because this $3\times 3$ rotation
is, in general, complex (unitary).  We want to avoid this complication
by all means.  }.

Still, there is some freedom remaining due to the fact that the
$\bar{E}$-base has not, as yet, been fixed.  Flavour rotation in the
$\bar{E}$-space can be used to remove some of the unphysical degrees
of freedom in ${\bf Y_L}\equiv\lambda_{0 ij}$ coupling\footnote{ If
the decomposition is unique, then all unphysical degrees of freedom
will be removed, because then the full U(3) rotation is absorbed into
$\bar{E}$ and every rotation in the $\Lx{\alpha}{}$-space will
``destroy'' some of the properties we want to keep.}.  As every
general, complex matrix, $\lambda_{0 ij}$ can be uniquely decomposed
(polar decomposition theorem~\cite{la}) into a product of positive
semi-definite Hermitian matrix $\hat\lambda_0$ and unitary matrix
$V_E$ :
\begin{equation} 
\lambda_{0 ij} \ = \ \hat{\lambda}_{0 ik}\:  (V_E)_{kj}\;,
\end{equation}
$V_E$ can be then adsorbed in the chiral superfield $\bar{E}$
redefinition and the `hat' over $\lambda$ is dropped.

After all the transformations described above, we arrived to the form
of the superpotential~(\ref{superpot1}) where ${\bf (Y_D)}_{ij} =
\lambda'_{0 ij}$ and ${\bf (Y_U)}_{ij}$ are flavour-diagonal and ${\bf
(Y_L)}_{ij}=\lambda_{0 ij}$ is hermitian.  Other coupling constants
are free and, in general, complex parameters.

\subsection{Block Diagonalising}
\label{sec:block}

In the following sections we outline the procedure by which the
parameters of the general hermitian matrix ${\bf (Y_L)}_{ij}$ can be
initialised such that the correct values are obtained for the charged
lepton masses and the MNS mixing matrix~\cite{MNS}.  In order to do
that, it will be convenient to diagonalise the neutralino-neutrino and
chargino-charged lepton mass matrices in two stages.  First an
approximate, unitary or biunitary transformation will result in
matrices in block diagonal form; the standard model and supersymmetric
fermion masses being split into separate blocks.  Then, a second
transformation will diagonalize the blocks.

The block diagonalisation can be performed for any complex matrix.
Every general matrix $M$ can be diagonalised by two unitary matrices
$V,U$:
\begin{equation} 
V^{\dagger} M U = \hat{M} = \textup{diag}(m_1,m_2,\cdots,m_n) = \left(
\begin{array}{cc} M_1 & 0 \\ 0 & M_2 \\ \end{array} 
\right) \;,
\end{equation}
where $m_i^2$ are eigenvalues of $MM^\dagger$ and $M_1, M_2$ are two
diagonal sub-matrices of a chosen size.  Hence, one can always rewrite
$M$ in the form
\begin{equation} 
M = V \hat{M} U^{\dagger} = V A^{\dagger} A \hat{M} B^{\dagger} B 
U^{\dagger} \;,
\end{equation}
where $A$ and $B$ are some unitary matrices of the form
\begin{equation} 
\begin{array}{cc} A= \left( \begin{array}{cc} A_1 & 0 \\ 0 & A_2 \\
\end{array} \right)\;,\qquad  B = \left( \begin{array}{cc} B_1 & 0 \\ 0 & B_2 \\
\end{array} \right) \end{array}  \;,
\end{equation}
with sub-matrices $A_{1,2},B_{1,2}$ which are also unitary.  Thus we
can write
\begin{equation} 
M = Q M_B P^{\dagger}  \;,
\label{eq:block}
\end{equation}
where
\begin{equation} 
M_B = A \hat{M} B^{\dagger} = \left( \begin{array}{cc} A_1 M_1
B_1^{\dagger} & 0 \\ 0 & A_2 M_2 B_2^{\dagger} \end{array} \right)\;,
\end{equation}
is block diagonal in form and $Q=VA{^\dagger}$, $P=UB^{\dagger}$.  Of
course, $M_B$ is not uniquely defined.

Block diagonalisation is particularly useful in case of hierarchical
matrices, when one can find analytical approximate formulae for $P, Q$
matrices in eq.(\ref{eq:block}).  Consider for example a hermitian
matrix (other cases can be considered analogously) of the form:
\bea
M = \left(\begin{array}{cc}
m_A & m_B \\
m_B^{\dagger} & m_C
\end{array}\right) \;,
\eea
where $m_A=m_A^{\dagger}$, $m_C=m_C^{\dagger}$ and $||m_A||\gg
||m_B||,||m_C||$.  In such case, the approximately (up to the terms
${\cal O}(||m_{B,C}||^2/||m_A||^2)$) unitary matrix $U$,
\bea
U = \left(\begin{array}{cc}
1 & - (m_A)^{-1}m_B \\
m_B^{\dagger}(m_A)^{-1} & 1
\end{array}\right)\;,
\eea
transforms $M$ into approximately block-diagonal form:
\bea
U^\dagger M U &=& \left(\begin{array}{cc}
m_A {+} (m_A)^{-1}m_B m_B^{\dagger} + m_B m_B^{\dagger}(m_A)^{-1} &
(m_A)^{-1} m_B m_C \\
m_C^{\dagger} m_B^{\dagger} (m_A)^{-1} & m_C - m_B^{\dagger}
(m_A)^{-1} m_B
\end{array}\right) + {\cal O} \left( ||m_{B,C}||^3 \over ||m_A||^2 \right) \nonumber \\[4mm]
&\approx&\left(\begin{array}{cc}
m_A & 0 \\
0 & m_C - m_B^{\dagger} (m_A)^{-1} m_B
\end{array}\right) + {\cal O}\left({||m_{B,C}||^2\over||m_A||}\right) \;.
\label{eq:appbl}
\eea
We kept explicitly the ${\cal
O}\left({||m_{B,C}||^2\over||m_A||}\right)$ term in (22) element of
block-diagonalised form of $M$ as in many models $m_C\equiv 0$ and in
this case it will be the only term which survives (the see-saw
mechanism~\cite{Schechter:1981cv,GORAN} for neutrino masses being the most famous
example of this hierarchical structure).

As a next step one needs to find matrices that diagonalise the
sub-blocks in eq.(\ref{eq:appbl}).  We employ this method in the next
sections where we present explicit perturbative ($1_\mathrm{st}$
order) results for analogs of the matrices $Q$ and $P$ in
Eq.~(\ref{eq:block}), in both neutral and charged fermion masses in the
\LMSSM.
In passing that, although we use only the approximate analytical
expressions for the see-saw type expansions above, in our numerical
predictions we perform exact block diagonalization, iteratively
finding the correct, and strictly unitary, matrices $P,Q$ of
eq.(\ref{eq:block}).

\subsection{Fermion masses and mixing}
In the following section we shall present the tree level phenomena of the
fermion sector in the \LMSSM.  We consider in turn, the neutral and charged
fermion sectors and the patterns of the mass matrices.  
We consider the tree level eigenvalues, particularly for the neutral sector and the approximate block 
diagonalisation of the matrices.  In section~\ref{subsec:mns}, we use the
approximate block diagonalisation to consider the way in which the MNS matrix appears in
this model, as it is now a sub-block of a larger unitary matrix, the MNS matrix itself is
not generally unitary.  This analysis is then used to ensure the correct 
low energy parameters are reproduced,
despite mixing between the leptons and heavy fermion fields.

In section~\ref{sec:3} we consider the effect of radiative corrections at the
order of one-loop.  After setting out the renormalisation framework, we
present in turn various loop diagrams and highlight the important
contributions.  The full numerical analysis has been completed, however
approximate expression are presented for each contribution which demonstrate
from where the important effects arise.  We will consider the case where the
tree-level effect dominates and gives rise to the larger, atmospheric mass
squared difference, in which case the solar mass squared difference is
generated by the loop effects.  We also show that it is possible for loop
effects to be greater than the tree level affects, in which case both mass
squared differences are generated at the level of one-loop.

\subsubsection{Neutral fermion sector}

In the lepton number violating extension of the minimal supersymmetric
standard model (\LMSSM) the neutrinos ($\nuL{1,2,3}{}$), neutral higgsinos 
($\nuL{0}{}$ and $\hinobz{}$)and
neutral colourless gauginos ($\Wino{0}{}$ and $\Bino{}{}$) mix.  
To transform the fields into the
mass basis, the $7\times7$ neutralino mass matrix must be
diagonalised.  In the interaction basis,
\begin{equation} 
\Lag \supset -\half \left( \begin{array}{cccc} - i \Bino{}{}, & -i
	\Wino{0}{}, & \hinobz{}, & \nuL{\alpha}{} \\ \end{array}
	\right) \mama{N} \left( \begin{array}{c} - i \Bino{}{} \\ -i
	\Wino{0}{} \\ \hinobz{} \\ \nuL{\beta}{} \\ \end{array}
	\right) \ + \ {\rm H.c}\;,
\label{eq:nut}
\end{equation}
where the full $7\times7$ mass matrix reads
\begin{equation} 
\mama{N} = \left( \begin{array}{cc} M_{N \, \tiny{4\times4}} & d_{N \,
	\tiny{4\times3}} \\
	d^T_{N \, \tiny{3\times4}}& 0_{\tiny{3\times3}}
\label{eq:mnsplit}
\end{array}  \right) \;,
\end{equation}
and the sub-blocks are, in the basis $(- i \Bino{}{}, -i \Wino{0}{},
\hinobz{}, \nuL{\alpha}{}\equiv \tilde h^0_1,\nu_i)$~\cite{DRRS}
\begin{equation}
M_{N \, \tiny{4\times4}}=\left( \begin{array}{cccc} 
M_1 & 0 & \frac{g v_u}{2} & - \frac{g v_d}{2} \\ 
0 & M_2 & - \frac{g_2 v_u}{2} & \frac{g_2 v_d}{2} \\ 
\frac{g v_u}{2} & - \frac{g_2 v_u}{2}& 0 & -\mu_0 \\ 
- \frac{g v_d}{2}& \frac{g_2 v_d}{2} & -\mu_0 & 0 \\
\end{array} \right)\;,
\end{equation}
and
\begin{equation} 
d_{N \tiny{4\times3}}= 
\left( \begin{array}{cccc}  
0 & 0 & 0 \\
0 & 0 & 0 \\
-\mu_1 & -\mu_2 & -\mu_3 \\
0 & 0 & 0\\ \end{array}  \right) \;.  
\end{equation}
There is no quantum number to differentiate between neutralinos and
neutrinos, the states of definite mass do not have definite lepton
number and, as such, there is no reason to think of neutrinos and
neutralinos separately.  However, for realistic values of parameters,
four of the mass eigenstates are heavy and three are very light, so it
is convenient to refer to them as to neutralinos and neutrinos,
respectively.  In addition to this, it can be seen that the mixing is
sufficiently small that these three light neutral states are the states which
dominantly appear in the decay of the W boson to charged leptons, 
differentiating between the eigenstates we refer to as neutrinos from those we
refer to as neutralinos.

The matrix,$\ZN{}$, which rotates the fields in (\ref{eq:nut}) from the
interaction basis to the mass eigenstate basis is given by
\begin{equation}
\left( \begin{array}{c} -i \Bino{}{} \\ -i \Wino{}{0} \\ \hinobz{} \\
	\nuL{\alpha}{} \end{array} \right) = \ZN{} \left( \begin{array}{c}
	\Kz{1}{} \\ \vdots \\ \Kz{7}{} \end{array} \right) \;.
\end{equation}
where $\Kz{1,\ldots,7}{}$ are seven neutral two component spinors.

The matrix $\mama{N}$, as it has been split in eq.(\ref{eq:mnsplit}),
contains block diagonal terms that conserve lepton number and off
diagonal blocks which violate lepton number.  The latter are expected
to be very small, as they are strongly constrained by the bounds on
neutrino masses or other lepton number violating processes.  Thus, one
can use the block diagonalization procedure of section~\ref{sec:block}
and, neglecting terms of the order $d_N^2\over M_{N}^2$ and assume
$\ZN{}$ to be of the form
\begin{equation} 
\ZN{} =\left( \begin{array}{cc} 
1 & - M_N^{-1} d_N \\ 
{d_N^{\dagger}} M_N^{{\dagger} -1} & 1
\end{array} \right) 
\left( \begin{array}{cc} 
{\cal Z}_N & 0\\ 
0 & {\cal Z}_{\nu}
\end{array} \right) \;.  
\label{eq:zndef}
\end{equation}
The first matrix on the RHS of eq.(\ref{eq:zndef}) which is the
analog of the matrix $Q^\dagger$ in (\ref{eq:block}) block
diagonalises the neutrino-neutralino mass matrix:
\bea
\left( \begin{array}{cc} 
1 & M_N^{{\dagger} -1} {{d_N^*}} \\ 
- d_N^T M_N^{-1} & 1
\end{array} \right)
\mama{N}
\left( \begin{array}{cc} 
1 & - M_N^{-1} d_N \\ 
{{d_N^\dagger}} M_N^{{\dagger} -1} 
& 1
\end{array} \right)
\approx \left( \begin{array}{cc} 
M_{N \, \tiny{4\times4}} & 0 \\ 
0 &m_{\nu \, \tiny{3\times3}}^{eff} 
\end{array} \right)\;,
\eea
where the ``TeV'' see-saw suppressed effective $3\times 3$ neutrino
mass matrix is given by~\cite{Joshipura,Banks,Nowakowski}
\bea
m^{eff}_{\nu} =  - d_N^T M_N^{-1} d_N
= {v_d^2(M_1 g_2^2 + M_2 g^2)\over 4 \textup{Det}[M_N]} \left(
\begin{array}{ccc} 
\mu_1^2& \mu_1\mu_2& \mu_1\mu_3\\
\mu_1\mu_2&\mu_2^2&\mu_2\mu_3\\
\mu_1\mu_3&\mu_2\mu_3&\mu_3^2\\
\end{array} \right) \;.
\label{eq:eff}
\eea
Physical neutralino masses and mixing matrix ${\cal Z}_N$ can be found
in a standard manner by numerical diagonalization of the matrix $M_N$.
Diagonalization on $m_{\nu}^{eff}$ can be easily done analytically,
leading to two massless and one massive neutrino, with its mass given
by:
\begin{equation}
m_{\nu}^{tree} =\left| {v_d^2(M_1 g_2^2 + M_2 g^2)\over 4
\textup{Det}[M_N]} \right| (|\mu_1|^2 + |\mu_2|^2 +|\mu_3|^2)\;,
\label{nmass}
\end{equation}
and the mixing matrix ${\cal Z}_\nu$ is
\begin{eqnarray}
{\cal Z}_\nu = \left( \begin{array}{ccc}
	\frac{|\mux{2}|}{\sqrt{|\mux{1}|^2+|\mux{2}|^2}} &
	\frac{|\mux{1}||\mux{3}|}{\sqrt{|\mux{1}|^2
	+|\mux{2}|^2}\sqrt{|\mux{1}|^2+|\mux{2}|^2+|\mux{3}|^2}} &
	\frac{|\mux{1}|}{\sqrt{|\mux{1}|^2+|\mux{2}|^2+|\mux{3}|^2}}
	\\
	\frac{-|\mux{2}|\mux{1}}{\mux{2}\sqrt{|\mux{1}|^2+|\mux{2}|^2}}
	& \frac{\mux{1}\muxs{2} |\mux{3}|}{|\mux{1}|\sqrt{|\mux{1}|^2
	+|\mux{2}|^2}\sqrt{|\mux{1}|^2+|\mux{2}|^2+|\mux{3}|^2}} &
	\frac{|\mux{1}|\muxs{2}}{\muxs{1}\sqrt{|\mux{1}|^2+|\mux{2}|^2+|\mux{3}|^2}}
	\\ 0& - \frac{\mux{1}|\mux{3}|\sqrt{|\mux{1}|^2+|\mux{2}|^2}}
	{\mux{3}|\mux{1}|\sqrt{|\mux{1}|^2+|\mux{2}|^2+|\mux{3}|^2}} &
	\frac{|\mux{1}|\muxs{3}}{\muxs{1}\sqrt{|\mux{1}|^2+|\mux{2}|^2+|\mux{3}|^2}}
	\\
\end{array} \right)
\left( \begin{array}{c|c} X_{2\times2} & 0 \\ \hline 0 & 1 \end{array}
\right) \nonumber \;, \\[2mm]
\label{eq:zv}
\end{eqnarray}
where $X_{2\times2}$ is an $SU(2)$ rotation.  At tree level, the five
massive eigenstates are unambiguously defined by diagonalising the
mass matrix.  The two massless eigenstates, due to the fact that they
are degenerate in mass, are not fully defined.  The eigenstates are
chosen to be orthogonal, but it is still possible to perform a
rotation on the eigenstates.  As such, statements about the lightest
neutrinos, $\nu_{1,2}$,
are basis dependent.  Because of this, the one loop contributions to
$\hat{{\cal M}}_{N\,pq}$ are also basis dependent.  By choosing a
different linear superposition of the tree level eigenstates, the
one-loop contributions to the $2\times 2$ sub-lock $\hat{{\cal
M}}_{N\,(5,6)(5,6)}$ referring to the massless neutrinos would be
redistributed between themselves.  This freedom of basis choice is
only present at tree level and is not physical.  Thus, we start from
$X_{2\times2}=1_{2\times2}$ and after calculating the radiative
corrections to the neutralino-neutrino mass matrix we adjust
$X_{2\times2}$ such that the off-diagonal one-loop contribution
$\delta M_{N\,56}$ is approximately zero (this can be done
iteratively).  As such the effect of rediagonalising the neutrino
sector after loop corrections are added is small.  As we discuss in
section~\ref{subsec:mns}, choosing the basis in this manner helps also
to define the lepton Yukawa couplings in terms of measured quantities
like lepton masses and the $U_{MNS}$ mixing matrix.

The result that two of the neutrino masses vanish at the tree level is
not the effect of the approximations
made~\cite{Joshipura,Banks,Nowakowski}.  The explicit calculation of
the secular equation for the full neutralino-neutrino mass matrix $\mama{N}$,
results in
\bea
\det(\mama{N} - \lambda) &=& -\lambda^2 \left[ \lambda\det(M_N {-
\lambda})\right.  \\
&-&\left.  (\mu_1^2 +\mu_2^2 +\mu_3^2)\left(\lambda(M_1 - \lambda)(M_2
- \lambda) +{g_2^2v_d^2\over {4}}(M_1 - \lambda) + {g^2 v_d^2\over
  {4}}(M_2 - \lambda)\right)\right]\;.\nonumber
\eea
Hence, $\mama{N}$ always has at least two zero modes.  This can be seen
directly, by noting that the final three columns of the $7\times7$ mass matrix
are proportional to each other.

Finally, the physical eigenstates of neutralinos and neutrinos are
approximately given by, respectively:
\begin{equation}
{\cal Z}_N \left( \begin{array}{c} \Kz{1}{} \\ \Kz{2}{} \\ \Kz{3}{} \\
\Kz{4}{} \end{array} \right) = \left( \begin{array}{c} -i \Bino{}{} \\
-i \Wino{0}{} \\ \hinobz{} \\ \nuL{0}{} \end{array} \right)
+ M_N^{-1} d_N \nuL{i}{} \;, 
\end{equation}
and
\begin{equation}
{\cal Z}_\nu\left( \begin{array}{c} \Kz{5}{} \\ \Kz{6}{} \\ \Kz{7}{}
\end{array}\right) 
= - d_N^{{\dagger}} M_N^{{\dagger} -1} \left( \begin{array}{c} -i
\Bino{}{} \\ -i \Wino{0}{} \\ \hinobz{} \\ \nuL{0}{} \end{array}
\right) + \nuL{i}{} \;, \label{eq:znnu}
\end{equation}
with $\nuL{0}{}=\tilde{h}_1^0$ is the down type Higgsino.  For a quick
view of definitions see also Appendix~\ref{app:mass}.

\subsubsection{Charged fermion sector}

In a similar fashion to the neutral sector, charged leptons, gauginos
and higgsinos mix.  The full $5\times 5$ chargino mass matrix in the
basis of Ref.~\cite{DRRS} is given by
\begin{equation}
\mama{C} = \left( \begin{array}{cc} M_{C\, \tiny{2\times 2}} & 0 \\
d_{C\, \tiny{3\times 2}} & m_{C\, \tiny{3\times 3}} \end{array}
\right)\;,
\end{equation}
with the lepton number conserving sub-locks  
\begin{equation}
M_{C\, \tiny{2\times 2}} = \left( \begin{array}{cc} M_2 & {g_2
v_{{u}}\over \sqrt{2}} \\ {g_2 v_d \over \sqrt{2}} & \mu_{{0}} \\ \end{array}
\right)\;, \qquad m_{C \,ij} = \frac{v_d}{\sqrt{2}} \lambda_{0 ij}\;,
\label{yuk}
\end{equation}
and the lepton number violating being
\begin{equation}
d_{C\, \tiny{3\times 2}} = \left( \begin{array}{cc}
    0 & \mu_1 \\
	0 & \mu_2 \\
	0 & \mu_3 \\ \end{array} \right) \;.
\label{dc}
\end{equation}
The rotation matrices which transform between interaction eigenstates
and mass eigenstates are given by
\begin{equation}
\left( \begin{array}{c} -i \Wino{}{+} \\ \hinobp{} \\ \eR{i}{}
\end{array} \right) = \Zp{} \left( \begin{array}{c} \Kp{1}{} \\ \vdots
\\ \Kp{5}{}
\label{zplus}
\end{array} \right)\;,
\end{equation}
\bea
\left( \begin{array}{c} -i \Wino{}{-} \\ \eL{\alpha}{}  \end{array}  \right)
= \Zms{} \left( \begin{array}{c} \Km{1}{} \\ \vdots \\ \Km{5}{}
\end{array} \right)   \;, \label{zminus}
\eea
and, as such, the mass matrix is diagonalised
\begin{equation} 
\hat{\mama{C}} = \Zmd{} \mama{C} \Zp{} \;, \label{mamc} 
\end{equation}
where the `hat' denotes that the matrix is diagonal.   

The matrices $\Zp{}$ and $\Zm{}$ can be determined by the requirement
that they should diagonalise the Hermitian matrices
$\mama{C}^{\dagger}\mama{C}$ and $\mama{C}\mama{C}^{\dagger}$,
respectively.  The off-diagonal blocks in the latter two combinations
are small comparing to the diagonal ones, so one can again use
block-diagonalising approximation of section~\ref{sec:block}.  Keeping
just the leading terms in $1/M_{C}$ expansion, one obtains
\bea
\Zm{} &\approx& \left( \begin{array}{cc} 1 & {-} M^{\dagger \, -1}_C
	d^{\dagger}_C \\ d_C M^{-1}_{C} & 1 \end{array} \right) \left(
	\begin{array}{cc} {\cal Z}_- & 0\\ 0 & {\cal Z}_{l^-} \\
	\end{array} \right)\;, \nonumber\\[3mm]
\Zp{} &\approx& \left( \begin{array}{cc} {\cal Z}_+ & 0\\ 0 & {\cal Z}_{l^+} \\ \end{array}
\right)\;.
\label{zcha}
\eea
Substitution of (\ref{zcha}) in (\ref{mamc}) results in the physical
effective mass matrix
\begin{equation} 
%
\hat{\mama{C}} = \left( \begin{array}{cc}
{\cal Z}_-^\dagger M_C {\cal Z}_+ +{\cal O}\left({d_C^2\over M_C}\right)
& {\cal O}\left({d_C m_C\over M_C}\right)
\\ 
{\cal O}\left({d_C^{{2}} \over M_C^{{1}}}\right)
& {\cal Z}_{l^-}^\dagger m_C {\cal Z}_{l^+} + {\cal O}\left({d_C^2 m_C\over M_C^2}\right)
\\ 
\end{array} \right)\approx 
\left( \begin{array}{cc}
{\cal Z}_-^\dagger M_C {\cal Z}_+ & 0 \\ 
0 & {\cal Z}_{l^-}^\dagger m_C {\cal Z}_{l^+}\\ 
\end{array} \right)\;.\label{defmc}
\end{equation}
Then the matrices ${\cal Z}_+,{\cal Z}_-$ can be again determined as
diagonalising matrices for the $M_C^{\dagger} M_C$, $M_C
M_C^{\dagger}$ products, with the additional requirement that physical
fermion masses are real and positive.  Matrix $m_C$ in our basis is
hermitian and as such ${\cal Z}_{l^+}={\cal Z}_{l^-}\equiv {\cal Z}_l$.
Furthermore, physical eigenstates of fermion fields are given by
\bea
{\cal Z}_+\left( \begin{array}{c} \Kp{1}{} \\ \Kp{2}{} \end{array} \right)
&\approx& \left( \begin{array}{c} -i \Wino{}{+} \\ \hinobp{}{}
\end{array} \right)\;, \nonumber\\[3mm]
{\cal Z}_{l} \left( \begin{array}{c} \Kp{3}{} \\ \Kp{4}{} \\ \Kp{5}{}
\end{array} \right) &\approx & \left( \begin{array}{c} \eR{1}{} \\
\eR{2}{} \\ \eR{3}{}
\end{array} \right)\;,  
\eea
and	
\bea
{\cal Z}_-^{*} \left( \begin{array}{c} \Km{1}{} \\ \Km{2}{} \end{array}
\right) &\approx& \left( \begin{array}{c} -i
\Wino{}{-} \\ \eL{0}{} \end{array} \right) + M_C^{-1 T} d_C^T \left(
\begin{array}{c} \eL{1}{} \\ \eL{2}{} \\ \eL{3}{}
\end{array} \right) \;,  \nonumber\\[3mm]
{\cal Z}_{l}^{*}\left( \begin{array}{c} \Km{3}{} \\ \Km{4}{} \\
\Km{5}{} \end{array} \right) &\approx& -d_C^{*} M_C^{-1*} 
\left( \begin{array}{c} -i \Wino{}{-} \\ \eL{0}{} \end{array} \right) 
+ \left( \begin{array}{c} \eL{1}{} \\ \eL{2}{} \\ \eL{3}{}
\end{array} \right)\;.\label{eq:zm}
\eea
For a quick view of the full charged fermion mass matrix
see Appendix~\ref{app:mass}.

\subsection{Constructing the MNS matrix}
\label{subsec:mns}

The lepton mixing matrices appear in the charged current gauge boson
vertex.  Whereas in the lepton number conserving case the $U_{\rm
MNS}$ matrix is a $3 \times 3$ matrix describing the mixing of three
charged leptons into three neutral leptons, the R-parity violating
case has the mixing of five charged fermions into seven neutral
fermions, of which the $U_{MNS}$ is a $3 \times 3$ sub-matrix, only
approximately unitary.  Thus,
\bea
{\cal L} \supset \frac{g_2}{\sqrt{2}}\:  W_\mu^+  \bar{\nu}_{Li}' \bar{\sigma}^\mu  e_{Li}' \: + \: {\rm H.c}
= \frac{g_2}{\sqrt{2}}\: W_\mu^+ \: \Kzb{p}{} \:  \bar{\sigma}^\mu \: (U_{\rm MNS})_{pq} 
\: \Km{q}{} \: + \: {\rm H.c} \;,
\label{Lmns}
\eea
where primes refer to interaction eigenstates, and the MNS matrix,
\begin{equation}
U_{\rm MNS} = {\cal Z}_{\nu}^{\dagger} {\cal Z}_{l}^{*} + {\cal
O}\left({d_c d_N\over M_C M_N}\right)\;,
\label{eq:mns}
\end{equation}
is defined in terms of the mixing matrices introduced in (\ref{eq:zv})
and below (\ref{defmc}).

As the first term in Eq.~(\ref{eq:mns}) is unitary, unitarity violation in
$U_{\rm MNS}$ is at most of the order of
${d_c d_N\over M_C M_N} \sim \frac{m_{\nu}^{\textup{\tiny{tree}}}
M_{\textup{\tiny{SUSY}}}}{M_{Z}^2}\tan^2\beta \sim 10^{-12} \tan^2\beta$,
which is well below sensitivity of current (or planned) experiments
determining the MNS matrix.


\subsection{Input parameters}
\label{subsec:inputpar}

The parameters characterising the light charged fermions are already
very well known; masses are measured with very good accuracy.  In
contrast to this, the neutrino sector is not, as yet, known with the
same precision.  There is however, information about the mass square
difference between neutrinos and the mixing between different
interaction states, and upcoming experiments should improve our
knowledge of neutrino parameters in the near future.  Furthermore,
supersymmetric fermions have not yet been discovered, and their masses
and couplings (those which are not determined by supersymmetric
structure of the model) are entirely unknown.  In the \LMSSM both
sectors mix, and thus the question of effective and convenient
parameterisation arises.  In this section, we will consider the
parameters in the Lagrangian which effect the tree level masses and mixing.
Later, we discuss parameters which affect the neutral fermions at the order of
one loop.

As the SUSY sector has not been measured directly, it is convenient to
take as an input the following set of Lagrangian parameters: $M_1$,
$M_2$, $\tan\beta$, $\mu\equiv\mu_0$.  With $\mu_i$ of the order of $\textup{MeV}$, 
corrections to the
supersymmetric sector from the light fermion sector are see-saw
suppressed and negligible.  Chargino and neutralino masses and
lepton-number conserving couplings are thus to a very good accuracy
determined by the above four parameters.  Reconstructing their values
from the actual experimental measurements has already been discussed
in the literature~\cite{KALZER}.

In a next step, neutrino masses can be parameterised at tree level by
setting the lepton-number violating parameters $\mu_i$, $i=1,2,3$.  In
the future, when the neutrino mass matrix is known to better accuracy,
it could become more convenient to reconstruct $\mu_i$ from the
experimental data - for that, the knowledge of radiative corrections
to the neutrino masses would be vital.

To initialize the Lagrangian parameters in the light charged lepton
sector, one needs to input the lepton masses, $m_e$, $m_\mu$, $m_\tau$
and the mixing matrix $U_{\rm MNS}$.  The lepton rotation matrix
${\cal Z}_{l}$ can be then calculated from (\ref{eq:mns})
\begin{equation}
{\cal Z}_{l} \approx  {\cal Z}_{\nu}^{*}\, U_{MNS}^{*}\;.
\label{eq:zl}
\end{equation}
As we have seen from (\ref{eq:zv}), the neutrino mixing matrix, ${\cal
Z}_\nu$ is defined at tree level up to a $U(2)$ rotation for a given set of $\mu_i$.  
The same matrix is then defined
completely at one-loop where all the neutrinos are no longer degenerate.  
Thus, a complete definition of ${\cal Z}_l$ requires a one-loop
corrected neutrino mixing matrix.  Then, the light charged fermion
mass matrix $m_C$ in (\ref{yuk}), which is hermitian and proportional
to the Yukawa matrix $\lambda_{0 ij} = \frac{\sqrt{2}}{v_d} m_{C
\,ij}$ is given by:
\bea
m_C = {\cal Z}_{l} \: {\rm diag}(m_e, m_\mu, m_\tau) \: {\cal
Z}_{l}^\dagger \;.
\label{eq:mcdef}
\eea
Eq.~(\ref{eq:mcdef}) holds under the assumption that one-loop
corrections to ${\cal Z}_l$ are small.  Otherwise, one needs to find
$m_C$ iteratively, such that physical (i.e. loop corrected) ${\cal
Z}_\nu$ and ${\cal Z}_l$ produce the correct experimentally measured
$U_{MNS}$ matrix of eq.(\ref{eq:zl}).
 
As we have repeatedly mentioned so far, it is important to notice that
the matrix $m_C$ is not diagonal.

\section{One-loop neutrino masses in \LMSSM}
\label{sec:3}

\subsection{Renormalization issues}

As we have already seen in the previous section, the presence of the
bilinear lepton number violating mass term in the superpotential,
$\mu_i$, triggers the mixing between neutrinos and neutralinos.
Diagonalization of the full $7\times 7$ neutralino mass matrix
generates four heavy ``neutralino'' masses and one ``neutrino'' mass
at tree level.  Furthermore, the two remaining neutrinos become
massive at the one loop level due to the presence of other
lepton-number violating couplings and masses.

Physical neutralino masses are defined as poles of the inverse
propagator.  The appropriate formula can be derived from (\ref{c6})
and the definition for the one-particle irreducible (1PI) self-energy
functions,
\begin{multicols}{2}
	\flushright
	\includegraphics[bb=109 669 222 695]{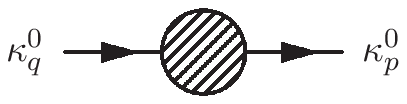}
\columnbreak
\flushleft
\begin{equation}
=  i \: \sigb{}{\mu} \: q_\mu \: \Sigma_{Npq}^L(q^2) \;,
\label{con1}
\end{equation}
\end{multicols}
\begin{multicols}{2}
	\flushright
	\includegraphics[bb=109 669 222 695]{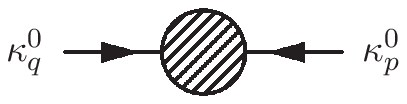}
\columnbreak
\flushleft
\begin{equation}
=- i \: \Sigma^D_{Npq}(q^2) \;,
\label{con2}
\end{equation}
\end{multicols}
\noindent where the momentum $q^\mu$ flows from left to right, and
$\sigma^\mu$ are the Pauli matrices\footnote{We use Weyl spinor
notation in our calculation.  The corresponding formulae for
Weyl-propagators and vertices are defined in Appendix~\ref{app:weyl}
and in~\cite{preSUSY}.}.  The requirement that the determinant of the
inverse propagator is zero, leads to the expression for the physical
neutralino mass matrix
\begin{equation}
	m_{Npq}^{\rm pole} = m_{Nq}^{\rm bare}(\mu_R)\delta_{pq}+
	\left [ 
	 \Re e \Sigma_{N \, pq}^{D}(m_{Np}^2) - 
	  m_{Np} \: \Sigma_{N \, pq}^{L}(m_{Np}^2)  \right ] \;,
	  \label{ren}
\end{equation}
where $\mu_R$ is the renormalization scale and $\Sigma_{N \,
qq}^{D,L}$ the 1PI contributions to the effective action defined
(\ref{con1},\ref{con2}).  $m_{Np}$ are the diagonal tree level
neutrino masses (they are zero for the two massless neutrinos).  Our
renormalization analysis is similar to the one in
Ref.~\cite{Hempfling}.  We have also studied an on-shell
renormalization analogous to the one in~\cite{Kniehl}.  In this
scheme, the physical mass formula is similar to (\ref{ren}).

Some additional remarks regarding the present are in order :

\begin{itemize} 

\item[a)] The two one-loop induced neutrino masses are perfectly
defined at one loop through Eq.~(\ref{ren}).  We have proven both
analytically and numerically that these masses are finite and
numerically that they are gauge independent at one-loop order.  This
result remains valid also when one takes into account mixing between
them.

\item[b)] The one-loop formula for neutralino masses (\ref{ren}),
receives, in addition to diagonal corrections, off diagonal ones,
$m_{Nqq}^{\rm pole} \rightarrow m_{Nqp}^{\rm pole}$.  Physical
neutralino and neutrino masses are then obtained from the
diagonalization of $m_{Nqp}^{\rm pole}$ as $\hat{m}_{N}^{\rm pole} =
(1+ \delta Z_N^\dagger) m_{N}^{\rm pole} (1+ \delta Z_N)$.  Then the
corrected mixing matrix $Z_N$ has to be replaced by $Z_N \rightarrow
(1+\delta Z_N) Z_N$ everywhere in our expressions for the self
energies.  However, the corrections to $Z_N$ matrix are of the order
$\delta Z_{Npq}\sim \delta m_{Npq}/(m_{Npp} - m_{Nqq})$ and are small
if the tree level masses are not degenerate. In our case this happens
only for the two massless neutrinos, so we include off-diagonal
corrections, shifting ${\cal Z}_\nu \rightarrow (1+ \delta {\cal
Z}_\nu) {\cal Z}_\nu$ where $\delta {\cal Z}_\nu$ has only the upper
$2\times 2$ block non-trivial\footnote{Possible exception is the case
when $\mu_i$ parameters are very small, so that one-loop corrections
to the neutrino masses are of the order of tree level neutrino mass or
bigger. In this case one needs to rediagonalise the full $3\times 3$
neutrino mass matrix. This is done numerically in section~\ref{sec:4}
when presenting our results for $\mu_i=0$.}.  As discussed previously
in section~\ref{sec:2}, this is actually necessary to fix the neutrino
basis. The resulting corrections to lighter neutrino masses are
formally two-loop, but numerically important and have to be taken into
account.  One should note that, as mentioned in the previous point,
one-loop corrections to the light neutrino mass sub-matrix are finite
- going beyond the approximation described above would require
performing formal renormalization on the neutral fermion mass matrix.
Finally, similar considerations apply to the case of the charged
fermion mixing matrix ${\cal Z}_l$ in (\ref{eq:zl}).

\item[c)] As we have already mentioned, in our neutral scalar
basis~\cite{DRRS} the sneutrino vevs are zero at tree level.  Non-zero
sneutrino vevs will appear in general at one-loop.  As a result, the
neutrino tree level mass in Eq.~(\ref{nmass}) should be corrected.
However, loop induced vev contributions {\it do not arise} for the
massless neutrinos--they are generated outside the $2\times 2$ light
neutrino mass matrix--which is the case we are interested in.

\item[d)] We choose $\mu_R=M_Z$ as renormalization scale in
(\ref{ren}) were we input the $\overline{DR}$ parameters for
$m_{Nq}^{\rm bare}(\mu_R)$ at tree level.  These parameters are taken
after diagonalising the full neutralino mass matrix in
Eq.~(\ref{eq:mnsplit}).

\item[e)] The infinities which arise in the calculation of the one
loop corrections, must be absorbed in parameters of the tree-level
Lagrangian.  It is possible to check that there are no infinities
which must be absorbed where the mass matrix contains zero entries.
The divergent parts of the integrals do not depend on the masses of
the particles in the loop integral and as such the infinities only
arise when a diagram exists in the interaction picture with only a
mass insertion on the fermion in the loop.  That is, the symmetry
which prevents a term existing in the classical Lagrangian also causes
the divergent part to cancel in the mass basis.  This guarantees that
it is always possible to absorb the infinite part of the integral in
the bare parameters of the classical Lagrangian.

\end{itemize}

Having considered the above points we find that for the calculation of the
one-loop corrections to the eigenstates which are massless at tree level, it
is sufficient to consider corrections to the bilinear terms purely between these
eigenstates.  We find that it is possible to neglect the one-loop effects which correct 
other entries of the neutral fermion mass matrix, describing neutralino masses
of neutralino-neutrino mixing.

\subsection{One loop contributions to the massless neutrino eigenstates}

The mixing of neutrino, neutral Higgsino and neutral gaugino
interaction eigenstates has been shown to result in two mass
eigenstates with zero mass.  It is important to note the composition
of the massless eigenstates; they consist solely of neutrino
interaction states, not containing any contribution from the fermionic
components of the gauge supermultiplets or the Higgs supermultiplets.
This can be stated, entirely equivalently as the
rotation matrix in Eq.~(\ref{eq:zndef}) becomes~\cite{Joshipura}
\begin{equation}
	\ZN{\{1\rightarrow4\}\{5\rightarrow 6\}}=0 \;.
\label{znrule}
\end{equation}
Radiative corrections at one-loop will affect all three of the light
mass eigenstates (`neutrinos') and will lift the degeneracy between
the massless eigenstates.  The possibility that the hierarchy of mass
differences in the neutrino sector can be explained in the \LMSSM is
considered.  If the `atmospheric' mass difference were to result from
the tree level splitting and the `solar' mass difference originated
from loop effects, the distinct hierarchy could be accommodated within
the model.  If the solar mass difference is to originate purely from
loop corrections to massless eigenstates, we must find loop
corrections from diagrams with external legs comprised purely of
neutrino interaction states.  A small caveat is required to compare
with the literature.  In a general basis where the sneutrino vacuum
expectation values are not zero, the massless neutrinos are comprised
of interaction state neutrinos and the interaction state Higgsino
which carries the same quantum numbers.  In the `mass insertion'-type
diagrams, this means that only diagrams without mass insertion or with
a mass insertion which changes the original neutrino external leg to
the down-type Higgsino can contribute to the solar mass~\cite{Abada},
if the assumption that the solar mass arises purely from loop
corrections to eigenstates which were massless at tree level.

The one-loop, one-particle irreducible self energies needed in
(\ref{ren}) are calculated in Appendix~\ref{app:weyl}, see
(\ref{c1},\ref{c2},\ref{c3},\ref{c4}).  Results are presented for
general vertices and for a general $R_\xi$ gauge.  One then has to
just replace these vertices with the appropriate Feynman rules of
Appendix~\ref{app:feyn} in order to obtain $\Sigma^{D,L}$.  Since
this, rather trivial replacement, leads to rather lengthy formulae for
the self energies, we refrain for presenting the full expressions
here.  Instead we examine in detail the dominant contributions
to the massless neutrinos, i.e., contributions to $\Sigma^D$.  Of
course, our numerical analysis exploits the full corrections.

From the expressions (\ref{c1},\ref{c2}), it can be seen that these
corrections are proportional to the mass of the fermion in the loop.
As such, the diagrams that give a large contribution are the diagrams
with sufficiently heavy fermions compared to any suppression from the
vertices.  In addition, standard model neutral fermion masses arise
entirely due to Supersymmetry breaking in the \LMSSM so corrections
are expected to be large for individual diagrams or a certain amount
of fine tuning is required for large SUSY soft breaking masses.

In the next section of this chapter we analyze analytically all the
possible contributions to $\Sigma^D_N$ for the massless neutrinos,
isolating the dominant ones.  For simplicity, we shall confine
ourselves only to the diagonal parts of $\Sigma^D_N$, although our
numerics account also for the off diagonal effects in the massless
neutrino sub-block.

\subsubsection{Neutral fermion - neutral scalar contribution}
\label{subsubsec:neutral}

Diagrammatically this contribution reads as:

\begin{center}
\hbox{\hspace*{50pt} \includegraphics[bb=109 640 242 713]{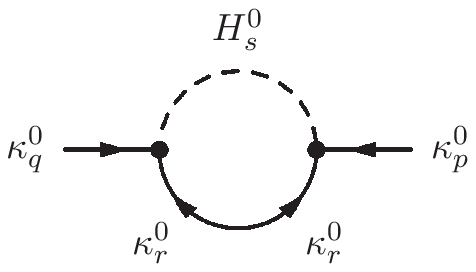}
\hspace{40pt}  \includegraphics[bb=109 640 242 713]{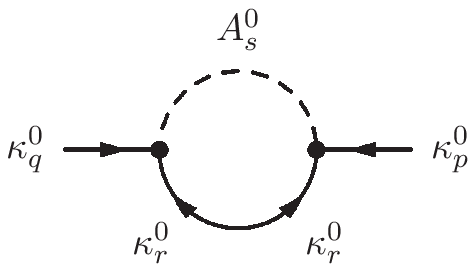}}
\end{center}

\noindent This can be easily calculated by using the formula
(\ref{c1}) and with the Feynman rules read from Appendix~\ref{app:feyn1}.
The result for the full contribution to the massless neutrinos,
$p=q=\{ 5,6 \}$, is :
\begin{eqnarray}
\Sigma^D_{N\, pp}& & = - \sum_{r=1}^{7} \sum_{s=1}^{5}
 \sum_{i,j=1}^{3} \frac{m_{\Kz{r}{}}}{(4\pi)^2} \times \nonumber
 \\[3mm] && \left [ \frac{e}{2 c_W} \ZN{(4+i)p} \ZN{1r} - \frac{e}{2
 s_W} \ZN{2r} \ZN{(4+i)p} \right ] \left [ \frac{e}{2 c_W} \ZN{(4+j)p}
 \ZN{1r} - \frac{e}{2 s_W} \ZN{2r} \ZN{(4+j)p} \right ] \nonumber
 \\[3mm] && \left [ \ZR{(2+i)s} \ZR{(2+j)s}
 B_0(m_{\Kz{p}{}}^2,m^2_{\Hz{s}},m^2_{\Kz{r}{}}) - \ZA{(2+i)s}
 \ZA{(2+j)s} B_0(m_{\Kz{p}{}}^2,m^2_{\Az{s}},m^2_{\Kz{r}{}}) \right ]
 \;,\label{fneutral}
\end{eqnarray}
\noindent where $\Hz{1,\ldots,5}$ and $\Az{1,\ldots,5}$ are the CP-even and
CP-odd neutral scalar fields, respectively, each containing a mixture of Higgs
and sneutrino fields.  The matrices $Z_N, Z_R, Z_A$ are those that diagonalize the
neutralino, CP-even, and CP-odd Higgs boson mass matrices and are defined
in (\ref{eq:zv}) and (\ref{zrdef},\ref{defza}), respectively (for
analytic expressions for $Z_R, Z_A$ see Eq.~(3.14) and (3.25) of
\cite{DRRS}).  Individually, the neutral fermion - neutral scalar
diagrams in (\ref{fneutral}) are large, however, if there were no
splitting between the mass of CP-even and CP-odd neutral scalar
eigenstates there would be an exact cancellation between the two
diagrams.  Notice also that the whole contribution is multiplied by a
neutralino mass which is generically of the order of the electroweak
scale.  It is rather instructive to simplify Eq.~(\ref{fneutral}) by
expanding around $m^2_{\Hz{s>2}}$ and $m^2_{\Az{s>2}}$ as,
\begin{eqnarray}
\Sigma^D_{N\, pp}& & \simeq -\sum_{r=1}^{7} \sum_{i=1}^{3}
 \frac{m_{\Kz{r}{}}^3}{4(4\pi)^2} {\cal Z}_{\nu \, ip}^2 \left [
 \frac{e}{c_W} {\cal Z}_{N\, 1r} - \frac{e}{s_W} {\cal Z}_{N\, 2r}
 \right ]^2 \frac{\Delta m_{\tilde{\nu} i}^2 }{(m_{\tilde{\nu} i}^2 -
 m_{\Kz{r}{}}^2 )^2} \ln\frac{m_{\Kz{r}{}}^2}{m_{\tilde{\nu} i}^2}\;,
\label{fneutral2}
\end{eqnarray}
where $\Delta m_{\tilde{\nu} i}^2 = m^2_{\tilde{\nu}_{+ i}} -
m^2_{\tilde{\nu}_{- i}}$ is the CP even - CP odd sneutrino square mass
difference.  Its analytical form can be derived from Eqs.~(3.14) and
(3.25) of~\cite{DRRS} to be
\begin{eqnarray}
\Delta m_{\tilde{\nu} i}^2 = \frac{B_i^2 \tan^2\beta}{M_A^2 - M_i^2} \
+ \ O(B_i^4/M_i^6) \;,
\label{snasn}
\end{eqnarray}
where $M_A$ is the CP-odd Higgs mass and $M_i$ the soft breaking
slepton masses which are diagonal in our basis, see Ref.~\cite{DRRS}.
A similar expression has been derived in Ref.~\cite{Haber}.
${\cal Z}_{\nu}$ and ${\cal Z}_N$ are defined in (\ref{eq:zndef}) and
(\ref{eq:zv}).  The contribution (\ref{fneutral2}) is driven by the
lepton number violating terms in the soft supersymmetry breaking
sector, $B_i$ and the whole expression for the neutral scalar
contribution collapses approximately to
\begin{eqnarray}
\Sigma^D \sim \left ( \frac{\alpha}{4\pi} \right ) \: m_{\Kz{}{}}
\left ( \frac{m_{\Kz{}{}}}{M} \right )^2 \: \frac{B_i^2 \:
\tan^2\beta}{(m_{\Kz{}{}}^2-M^2)^2} \; ,\label{nsn}
\end{eqnarray}
where $M$ is the sneutrino or Higgs and $m_{\Kz{}{}}$ the neutralino
masses in the loop, respectively.  The importance of this contribution
has already been pointed out in Refs.~\cite{Abada,Grossman}.  The mass
insertion approximation diagram reads as
\begin{center}
	\includegraphics[bb=108 637 243 711]{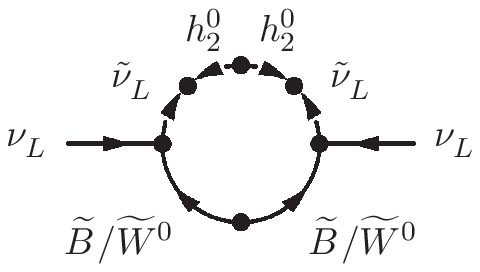}
\end{center}
where the `blobs' indicate insertions of $B$s, scalar and gaugino
masses.  The neutral scalar-fermion contribution is thus {\it a)}
suppressed from the CP-even-CP odd sneutrino mass square difference,
i.e, the lepton number violating soft SUSY breaking parameter $B_i$,
{\it b)} is enhanced by $\tan^2\beta$, and finally {\it c)} suppressed
by three powers of SUSY breaking masses.

The approximate formula (\ref{fneutral2}) does not in general capture
the full neutral fermion-scalar correction. There are other
corrections of the same order of magnitude, including the Higgs bosons
in $s=1,2$ states. This expansion is more complicated than
(\ref{fneutral2}) and is given explicitly in section~\ref{sec:4},
Eq.~(\ref{fneutralnew}), where we discuss our numerical results and
compare with approximate formulae of this chapter.

Ignoring other than the above diagrams possible cancellations, $[B_i
\tan\beta]$ must be smaller than the $0.1\%$ of the sneutrino mass
squared, $M^2$, in order to have $m_\nu \le 1 ~{\rm eV}$.  On the other hand,
numerically, if the `solar' neutrino mass difference were to be
generated by this diagram, then $B_i \sim {\cal{O}}(1)
\textup{GeV}^2$.  Because $B_i$ is in principle not constrained from
above by other means, we conclude that this diagram dominates the
whole contribution especially when the trilinear couplings,
$\lambda,\lambda'$, are negligible.

\subsubsection{Charged fermion - charged scalar  contribution}
\label{subsubsec:charged}

This contribution reads diagrammatically  as:
	\begin{center}
		\includegraphics[bb=109 640 242 713]{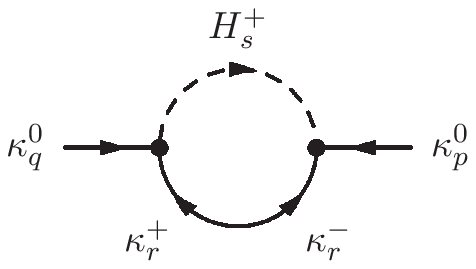}
\end{center}
Using the generic formula the self energy, (\ref{c1}), the Feynman
rules from Appendix~\ref{app:feyn2}, and also by applying
Eq.~(\ref{znrule}), we find that the full diagonal contribution for
$p=q= \{5,6\}$, is :
\begin{eqnarray}
\Sigma^D_{N\, pp} &=& \sum_{s=1}^{8} \sum_{r=1}^{5}
 \sum_{i,j,k,l=1}^{3} \sum_{\alpha, \beta =0}^3 \frac{m_{\Km{r}{}}
 }{(4\pi)^2} \left (\lam{\alpha l k} \ZHs{(2+\alpha)s} \Zp{(2+k)r}
 \ZN{(4+l)p} \right ) \times \nonumber \\[3mm]
&&\left [ \frac{e}{s_w} \ZH{(2+i)s} \Zms{1r} \ZN{(4+i)p} -\lam{\beta i
 j} \ZH{(5+j)s} \Zms{(2+\beta)r} \ZN{(4+i)p} \right ]
 B_0(m^2_{\Kz{p}{}},m^2_{\Hp{s}},m^2_{\Km{r}{}}) \;, \nonumber \\[2mm]
 \label{fcharged}
\end{eqnarray}
where $\ZN{}$, $\Zp{}, \Zm{}$, $\ZH{}$ are rotation matrices in the
neutral fermion, charged fermion, and charged scalar sectors, and
defined in (\ref{eq:zndef}), (\ref{mamc}), and (\ref{defzhp}),
respectively.  It is important to notice that following (\ref{mamc})
we obtain, $\Zp{(2+k)r} \simeq {\cal Z}_{l\, kr}$, with $r > 2$ and
hence the contribution (\ref{fcharged}) is proportional to the mass of
a light fermion, $m_{\Km{r}{}}$ .  In addition, since $\ZN{(4+l)p}
\simeq {\cal Z}_{\nu\, lp}$, (\ref{eq:zndef}) shows that the
contribution (\ref{fcharged}) contains the rotation mixing matrix
${\cal Z}_\nu$, which has been presented analytically in
(\ref{eq:zv}).  In order to analyze the dominant pieces from the
charged scalar - fermion contribution, it is instructive to consider
two cases : $\lam{ijk} = 0$ and $\lam{ijk} \ne 0$.

In the case where the trilinear superpotential couplings are absent
the charged lepton loop has a small contribution to the massless
neutrino eigenstates.  From the discussion above and (\ref{fcharged}),
we obtain at the limit of small lepton masses (compared to the SUSY
breaking ones),
\begin{eqnarray}
\Sigma^D_{N\, pp} &=&\sum_{s=1}^{8} \sum_{i,j,k,l=1}^{3}
  \sum_{r>2}^{5} \frac{m_{\Km{r}{}} }{(4\pi)^2} \left(\lam{0 l k} \:
  \ZHs{2s} \: {\cal Z}_{l \, kr} \: {\cal Z}_{\nu \, lp} \right)
  \times \nonumber \\[3mm]
&& \left [ \frac{e}{s_W} \ZH{(2+i)s} \Zms{1r} {\cal Z}_{\nu \, ip}
  -\lam{0 i j} \ZH{(5+j)s} \Zms{2r} {\cal Z}_{\nu \, ip} \right ] \:
  B_0(0, m^2_{\Hp{s}}, 0) \;,
  \label{acharged}
\end{eqnarray}
where $\lambda_{i0j}=-\lambda_{0ij}$ is the lepton Yukawa coupling
obtained from Eq.~(\ref{yuk}).  We can analyse further equation
(\ref{acharged}) by Taylor expansion with respect to $m^2_{\Hp{s}}$
(commonly named ``Mass Insertion Approximation'', or MIA, see, for example,
review in Ref.~\cite{Misiak:1997ei}) and using
(\ref{dc},\ref{mamc}) and (\ref{defzhp},\ref{mat:ch}) ,
\begin{eqnarray}
\Zm{1r} & \simeq & \left ( \frac{d_C^\dagger}{M_C} {\cal Z}_l \right
)_{1r} \simeq \frac{\mu_i}{M_C} \;, \\[2mm]
\ZHs{2s}\ZH{(2+i)s} m^2_{\Hp{s}} & = & {\cal M}^2_{H^+ \, 2,2+i}
\simeq \mu_i m_l \tan\beta \;, \\[2mm]
\ZHs{2s}\ZH{(5+i)s} m^2_{\Hp{s}} &=& {\cal M}^2_{H^+ \, 2, 5+i} \simeq
B_i \tan\beta \;,
\end{eqnarray}
where $m_l$ is the lepton mass and $M_C$ a generic gaugino mass.
Hence, the neutrino couples to either the right handed component of
the electron, the $\Wino{}{-}$ or the Higgsino with couplings
proportional to $\mu_i$.  All the above can be diagrammatically
depicted with mass insertions as :
\begin{multicols}{2}
\center
\includegraphics[bb=108 640 243 705]{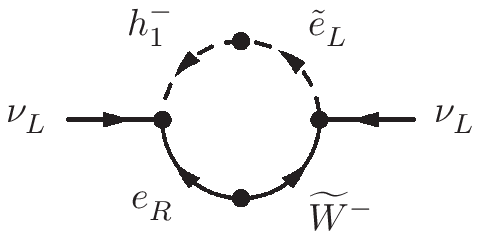}
\begin{equation}
\sim  m_l \frac{\mu_i}{M_C} \frac{B_i}{M_{H^+}^2} \tan\beta  \;,
\label{ch1}
\end{equation}

\columnbreak
\includegraphics[bb=108 638 243 705]{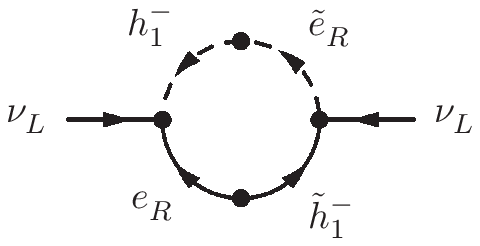}
\begin{equation}
\sim m_l \frac{\mu_i}{M_C} \frac{\mu_i m_l}{M_{H^+}^2} \tan\beta \;,
\label{ch2}
\end{equation}
\end{multicols}
\noindent  
where $M_{H^+}^2$ is a generic charged Higgs mass.  Obviously, both
the fermion and the scalar propagator are suppressed by lepton number
violating couplings.  This contribution is then compared with the
one previously considered with neutral particles in the loop.  Indeed,
in order to account for the atmospheric neutrino mass scale it must be
$\mu_i << \sqrt{B_i}\sim O(1 {\rm ~GeV})$ and hence the charged
particle contribution is {\it always} smaller than the neutral one.
Our finding here is in general agreement with the discussion in
Refs.~\cite{Chun,Grossman}.  Finally, notice the Goldstone contribution
vanishes since this always conserves lepton number.

If the trilinear superpotential lepton number violating coupling
$\lambda_{ijk}$ is turned on, then a lepton - slepton loop
contribution is generated.  In contrast with the pure bilinear case,
the trilinear contribution may dominate depending on the magnitude of
$\lambda$.  In this case the full contribution in (\ref{fcharged})
results in
\begin{eqnarray}
\Sigma^D_{N\, pp} &=& \sum_{s=1}^{8} \sum_{r=1}^{5}
 \sum_{i,j,k,l,m,n=1}^{3} \frac{m_{\Km{r}{}} }{(4\pi)^2} \times \left
 [ \lam{mik} \ZHs{(2+m)s} \Zp{(2+k)r} \ZN{(4+i)p} \right ] \times
 \nonumber \\[3mm]
&& \left [ \lam{lnj} \ZH{(5+j)s} \Zms{(2+n)r} \ZN{(4+l)p} \right ]
B_0(m^2_{\Kz{p}{}},m^2_{\Hp{s}},m^2_{\Km{r}{}}) \;.  \label{lamcon}
\end{eqnarray}
Again, making use of (\ref{eq:zv}, \ref{mamc}) we see that the
contribution is proportional to the light lepton masses and involves
the neutrino mixing matrix.  We can go a little bit further and
perform MIA expansion of~(\ref{lamcon}) as we did before.  The
contribution then reads,
\begin{eqnarray}
\Sigma^D_{N\, pp} &=& \sum_{i,j,k,l,m,n=1}^{3}
\frac{m_{l_q}}{(4\pi)^2} \lam{mik} \lam{lnj} {\cal Z}_{\nu \, ip}
{\cal Z}_{\nu \, lp} {\cal Z}_{l \, kq} {\cal Z}^*_{l \, nq} \times
\nonumber \\[3mm]
&& \left [ \frac{ ({\cal M}_{H^+}^2)_{2+m,5+j} }{ (\hat{\cal
M}_{H^+}^2)_{2+m} - (\hat{\cal M}_{H^+}^2)_{5+j} } \ln \frac{
(\hat{\cal M}_{H^+}^2)_{2+m} } { (\hat{\cal M}_{H^+}^2)_{5+j} } \right
] \;, \label{alamcon}
 \end{eqnarray}
where $m_{l_q}$ is a light charged lepton mass, $({\cal M}_{H^+}^2)$
is the charged scalar mass matrix in the interaction basis and is
given by (\ref{mat:ch}).  In our notation $(\hat{\cal
M}_{H^+}^2)_{5+j} \equiv (\hat{\cal M}_{H^+}^2)_{5+j,5+j}$, and so on.
In the denominator and logarithm of (\ref{alamcon}) one has the
difference of diagonal elements $mm$ and $jj$ of LL and RR slepton
mass matrices, respectively.  The approximation (\ref{alamcon}) is
proportional to the mixing matrix elements $({\cal
M}_{H^+}^2)_{2+m,5+j}$ which is nothing other than the LR mixing
elements of the charged slepton mass matrix (\ref{mat:ch}).  These
matrix elements are (almost) unbounded from experiments when $m=j$ in
contrast to the case $m\ne j$.

This contribution has been discussed largely in the literature, see
for instance~\cite{Haber,Grossman,Sacha,deCarlos,Borzumati,Chun,Choi}.
It is instructive to draw the mass insertion approximation diagram
corresponding to (\ref{alamcon}) :
\begin{center}
\includegraphics[bb=104 643 247 706]{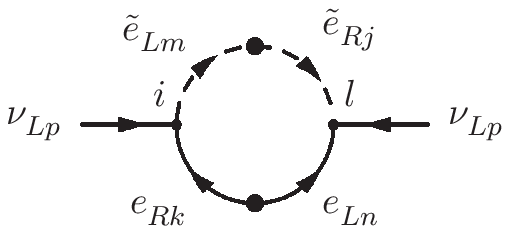}
\end{center}
In the case of dominant $\lam{ikk}$ coupling,
\begin{eqnarray}
\Sigma^D_N \sim \lambda^2 m_l^2 \: \frac{ \mu_0 \tan\beta + A_l}{M^2}
\;,\label{lepslep}
\end{eqnarray}
with $A_l$ being a trilinear SUSY breaking coupling and $M$ a generic
soft SUSY breaking mass for a slepton.  Comparing (\ref{lepslep}) with
(\ref{ch1},\ref{ch2}) of the previous case with $\lam{}\to 0$, we see
that the latter is suppressed with at least a factor $\mu_i/M$.  In
the case where the final two indices are different, $\lam{ikl}, k\ne
l$ there is an extra suppression from slepton intergenerational mixing
and the couplings must be stronger if the lepton in the loop is
lighter.  Our calculation is general enough to allow for these effects
too.  Furthermore, it is obvious from (\ref{alamcon}) that the $\tau -
\tilde{\tau}$-contribution, $\lam{i33}$, is the dominant one and this
coupling tends usually to be strongly bounded.

\subsubsection{Quark - squark contribution }
\label{subsubsec:quark}

In general, this contribution originates from up and down quarks and
squarks in the loop:
\begin{center}
	\hbox{ \hspace{50pt}
	\includegraphics[bb=109 643 242 712]{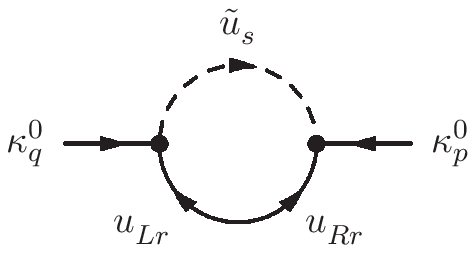} \hspace{40pt}
	\includegraphics[bb=109 641 242 715]{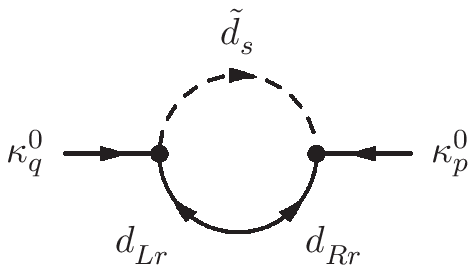}}

\end{center}

The up-quark-squark contribution vanishes identically, for the mass
eigenstates which are massless at tree level.  This can be
easily seen by applying the master equation (\ref{znrule}) to the
corresponding [neutralino-up-quark-up-squark] vertex given explicitly
in the Appendix~\ref{app:feyn3}.

The case of down quark-squark contribution (the right Feynman diagram
above) can be divided in two cases depending on the dominance of the
trilinear superpotential contribution : If $\lamp{} \to 0$ and the
only source of lepton number violation is the bilinear term then the
contribution vanishes.  Note that this does not necessarily disagree
with the findings of Refs.~\cite{Valle,Kaplan} where apparently this
contribution is claimed to be the dominant one.  Recall that we are
working in the basis of~\cite{DRRS} where the sneutrino vevs are zero
and thus we cannot directly compare, at least graph by graph with this
work.  In the case of Ref.~\cite{Kaplan} for example, the bilinear
term, $\mu_i L_i H_2$ is rotated away.  This rotation generates new,
non-negligible superpotential trilinear couplings which is the case we
are about to consider.  Hence, if $\lamp{ijk}\ne 0$, then the
situation changes dramatically.  Following (\ref{c1}), the Feynman
rules for the down type quarks of the Appendix~\ref{app:feyn3}, we
find that the most general contribution to the massless neutrinos,
$p=q=\{5,6\}$, reads as,
\begin{eqnarray}
\Sigma^D_{N\, pp} = \sum_{i,j,k,n,m=1}^3 \sum_{s=1}^6 \frac{3
m_{d_k}}{(4\pi)^2} \left [ \lamp{jik} \lamp{nkm} \Zsds{is}
\Zsd{(3+m)s} \ZN{(4+j)p} \ZN{(4+n)p} \right ] B_0(m^2_{\Kz{p}{}},
m_{\tilde{d}_s}^2, m_{d_k}^2) \;, \nonumber \\[2mm]
\label{qsq}
\end{eqnarray}
where the rotation matrix in the down squark sector, $\Zsd{}$, is
defined in (\ref{sdown}), and $Z_N$ in (\ref{eq:zndef}).  It is much
more instructive to Taylor expand the full contribution (\ref{qsq})
around a constant SUSY breaking mass into parameters of the original
Lagrangian.  In the limit of small neutrino and quark masses, this
results in
\begin{eqnarray}
\Sigma^D_{N\, pp} = \sum_{j,n,k,m=1}^3 \frac{{\cal Z}_{\nu \, j p} \:
{\cal Z}_{\nu \, np}}{(4\pi)^2} \: \left [ 3\: m_{d_k} \lamp{jik}
\lamp{nkm} \frac{ ({\cal M}^2_{\tilde{d}})_{i,3+m} }{(\hat{\cal
M}^2_{\tilde{d}})_{i} -(\hat{\cal M}^2_{\tilde{d}})_{3+m}} \ln
\frac{(\hat{\cal M}^2_{\tilde{d}})_{i}}{(\hat{\cal
M}^2_{\tilde{d}})_{3+m}} \right ] \;, \label{aqsq}
\end{eqnarray}
where the mass matrix ${\cal M}^2_{\tilde{d}}$ is defined in
(\ref{matd}).  Notice that $({\cal M}^2_{\tilde{d}})_{i,3+m}$ are the
elements of the LR mixing block of ${\cal M}^2_{\tilde{d}}$ and our
notation reads $(\hat{\cal {M}}^2_{\tilde{d}})_{i} \equiv ({\cal
M}^2_{\tilde{d}})_{ii}$.  The Feynman diagram with quark and squark
mass insertions representing (\ref{aqsq}) is:
\begin{center}
	\includegraphics[bb=104 640 248 708]{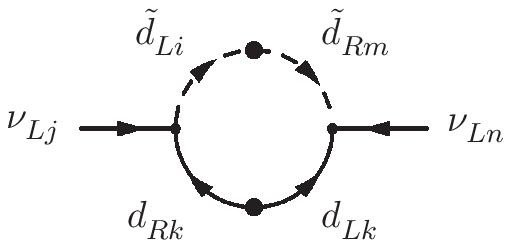}
\end{center}
Some remarks are in order : First the quark-squark contribution is
proportional to neutrino mixing through the matrix (\ref{eq:zv}), and
hence to possible hierarchies between $\mu_i$s.  Second, it is
proportional to squark flavour mixing.  Experimental results for
$K-\bar{K}$, and $B-\bar{B}$ mass difference set severe constraints in
the intergenerational squark mixings in the lepton number conserving
MSSM [$({\cal M}^2_{\tilde{d}})_{i,3+m}$ must be small for $i\ne m$].
Although, our calculation is as general as possible and allows for
these effects we shall assume $({\cal M}^2_{\tilde{d}})_{i,3+m}=0,
i\ne m$ in our numerical results below.  The quark-squark contribution
may be dominant for sufficient large $\lambda'$ couplings.


\subsubsection{ Neutral fermion - Z gauge boson contribution }
\label{subsubsec:z}

The corresponding Feynman diagram is :
\begin{center}
	\includegraphics[bb=109 640 242 711]{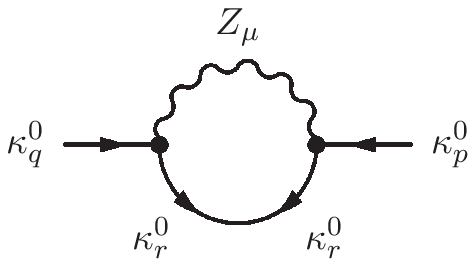}
\end{center}
Due to the approximate unitarity of the neutrino sub-block of
$\ZN{}{}$, the contribution of this diagram is suppressed either by
the lightness of the particle in the loop or by the value of the
coupling.  However, as we obtain from Eq.~(\ref{c2}) and
Appendix~\ref{app:weyl}, this contribution is gauge dependent.  The
dependence again cancels the neutral fermion-scalar contribution in
(\ref{fneutral}) with the Goldstone boson $(s=1)$ in the loop.
Although we prove this cancellation numerically, it can be also shown
analytically.


\subsubsection{Charged fermion - W gauge boson  contribution} 
\label{subsubsec:w}

The Feynman diagram  for this contribution is:  
\begin{center}
	\includegraphics[bb=109 640 242 712]{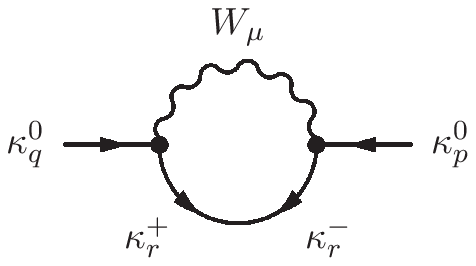}
\end{center}

\noindent Following (\ref{znrule}) and the Feynman rules of
Appendix~\ref{app:feyn4}, when the external legs are purely neutrino
interaction eigenstates $(p,r=5 {\rm ~or~} 6)$, there is no
$\Kz{}{}$-$\Kp{}{}$ vertex.  Hence, the contribution of this diagram
vanishes identically.

\subsubsection{Summary of the one-loop radiative corrections to
massless neutrinos }

The total one loop contribution to massless neutrino masses is given
by the sum of the neutral scalar loop in (\ref{fneutral}), the charged
scalar loop in (\ref{fcharged}), and the squark loop in (\ref{qsq}).
The gauge boson contributions are negligible.  If the trilinear
superpotential couplings are tiny then the dominant contribution
arises from the neutral scalar fermion loop and is proportional to
CP-even -- CP-odd sneutrino mixing [see Eq.~(\ref{fneutral2})].  If
trilinear couplings are not small, then depending upon their nature
$\lam{}$ or $\lamp{}$ dominate through $lepton-slepton$ [see
Eq.~(\ref{lamcon})] and $quark-squark$ [see Eq.~(\ref{aqsq})] diagrams.

\subsection{Comparison with Literature}

Our work improves on other work which can be found in the literature
as no assumptions or approximations need to be made.  Calculations can
be performed in the most general supersymmetric model with minimal
particle content, without any assumption that matrices are flavour
diagonal, or that any complex phases are set to zero.  We have not
neglected any terms or phases in the neutral scalar
sector~\cite{DRRS}, a basis was chosen in which to perform the
calculation that had a decoupled CP-odd and CP-even sector and two
real vevs.  In choosing this basis, it is clear that the lepton Yukawa
matrix is not, in general, diagonal and the lepton mixing matrix does
not come purely from the neutrino sector.  This is in contrast to
previous work where, in whatever basis the calculation is performed,
the lepton Yukawa is chosen to be diagonal.  In \cite{ValleD68}
assumptions are made in the soft sector, such as intergenerational
mixing being zero, which allows a basis to be chosen where the Yukawa
matrices are diagonal.  Similarly, in \cite{Haber,Grossman} there is
the assumption of CP conservation in the neutral scalar sector.

Many diagrams are suggested in the literature as being important in
generating a correct solar mass difference.  Under the assumption that
the solar mass difference comes solely from loop corrections to
eigenstates which are massless at tree level, in a general basis the
external legs must consist purely of neutrino and down-type Higgsino
interaction eigenstates (in the basis with sneutrino vevs rotated to
zero the external legs must consist purely of interaction state
neutrinos).  As such, when diagrams are presented with `mass
insertions', it is clear that any diagrams with insertions coupling
the neutrino to an up-type Higgsino or gaugino on the external leg
will not contribute to the solar mass difference at one loop.  In a
basis where sneutrino vevs are not zero, the diagrams with an
insertion mixing between interaction state neutrinos and the down-type
Higgsino contribute to the radiative correction of massless tree level
eigenstates.  In the basis where sneutrino vevs are zero this
contribution is included in the trilinear vertex, $\lambda^{(')}$.

Many papers~\cite{Haber,Grossman,Sacha,deCarlos,Chun} note the
contribution of the loops driven by trilinear couplings
$\lambda^{(')}$ and produce expressions, often with flavour mixing
suppressed, that agree with the expressions given here.

The contribution to the charged scalar loop from bilinear couplings is
also widely noted.  Whether a contribution is due to bilinear or
trilinear couplings is a basis dependent statement \cite{Sacha}.  We
agree with the results in \cite{ValleD68,Grossman,Kaplan}, however in
our basis the diagrams in \cite{ValleD68,Kaplan} are accounted for in
the trilinear loops.

The importance of the neutral scalar loop has also been noted
previously.  We agree with the general result of \cite{Hirsch:1997vz,GHsneutrino}
that a sneutrino mass difference will give rise to a radiative
correction in the neutrino sector and with \cite{Grossman} that this
loop can be the dominant contribution. The neutral scalar
contribution is included in the analysis presented in~\cite{Valle}, but
is not discussed in~\cite{ValleD68}. 

The role of tadpole corrections is stressed in \cite{Valle}.  If we
assume the solar mass difference arises from the loop corrected
`massless' neutrinos, we can see that the tadpoles do not play a role
the determining its magnitude.  In the interaction picture, there is
no $\nu_\alpha$-$\nu_\alpha$-Higgs vertex, so the tadpole
contributions vanish.  Of course, the tadpoles will affect the other
heavy neutral fermions.

A certain class of two-loop diagrams and resulting effects on bounds for 
lepton number violating couplings have been considered~\cite{Borzumati:2002bf}

\section{Numerical Results}
\label{sec:4}

In this section we present our numerical results for the neutrino
masses. As we have already explained, in our most general analysis we
use the MNS matrix defined by neutrino oscillations as an input.  Of
course this matrix is not accurately known, but its general `picture'
has been emerging during the last five or so years with angles and the
3$\sigma$ allowed ranges of the neutrino oscillation parameters from a
combined, global data, analysis~\cite{Schwetz}, reading,
\begin{eqnarray}
\sin^2\theta_{12} =0.24-0.40 \;, \qquad \sin^2\theta_{23} =0.34-0.68
\;, \qquad \sin^2\theta_{13} \le 0.046 \;, \label{nudata1} \\[3mm]
\Delta m_{21}^2 =(7.1-8.9)\times 10^{-5} \; {\rm eV}^2\;, \qquad
|\Delta m_{31}^2| =(1.4-3.3)\times 10^{-3}\; {\rm eV}^2\;.
\label{nudata2}
\end{eqnarray}
In our analysis we fix the neutrino mixing angles to reproduce the
tri-bimaximal mixing scenario of Ref.~\cite{Perkins} ,
\begin{eqnarray}
\sin^2\theta_{12} =\frac{1}{{3}} \;, \qquad 
\sin^2\theta_{23} =\frac{1}{{2}} \;, \qquad  \sin^2\theta_{13} = 0  \;,
\label{input}
\end{eqnarray}
in agreement with (\ref{nudata1}); the resulting predictions for
neutrino mass squared differences are then compared with
(\ref{nudata2}), to see whether the values chosen for the input
parameters give results in agreement with current experimental
limits. At present, there is no experimental evidence for CP-violation
in the leptonic sector; as such, although our analysis is general
enough to accommodate these effects, in what follows, we shall assume
that they are negligible.

In addition to the experimental inputs for the quark and lepton
fermion masses and mixings, soft supersymmetry breaking masses and
couplings must also be initialised. We follow the benchmark
SPS1a~\cite{SPS} where
 \begin{eqnarray}
 M_0=100~{\rm GeV}\;, \quad 
 M_{1/2}=250~{\rm GeV}\;, \quad 
 A_0=-100~{\rm GeV}\;, \quad 
 \tan\beta =10 \;,\quad
 \mu_0 > 0 \;, \nonumber \\
 \end{eqnarray}
and read the low energy SUSY breaking and superpotential parameters at
low energies using the code of~Ref.~\cite{Ben}.  The input parameters
of primary interest are those which violate lepton number. In the
basis of~\cite{DRRS}, these are,
\begin{eqnarray}
\mu_i \;, \qquad 
B_i \;, \qquad
\lambda_{ijk}  \;, \qquad 
\lambda'_{ijk} \;, \qquad
h_{ijk} \;, \qquad
h'_{ijk} \;, \label{lnvp}
\end{eqnarray}
where the last two, $h$ and $h'$ are the trilinear lepton number
violating parameters in the supersymmetry breaking part of the
Lagrangian.  Apart from these latter parameters, which concern
trilinear couplings of scalar particles, all others can be used to set
the atmospheric neutrino mass$^2$ difference or the solar mass$^2$
difference. There are two main cases:
\begin{itemize}  
\item Tree level dominance : the atmospheric mass$^2$ difference
originates from tree level contributions to neutrino masses
(Fig.~\ref{figlevel1}).
\begin{figure}[h] 
   \centering
   \includegraphics[width=5in]{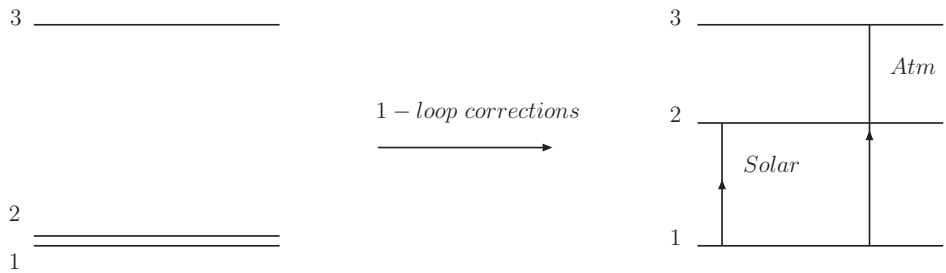} 
   \caption{Neutrino mass scales : tree level dominance}
   \label{figlevel1}
\end{figure}
\item Loop level dominance : The atmospheric mass$^2$ difference
originates from one-loop contributions to neutrino masses
(Fig.~\ref{figlevel2}).
\begin{figure}[h] 
   \centering
   \includegraphics[width=5in]{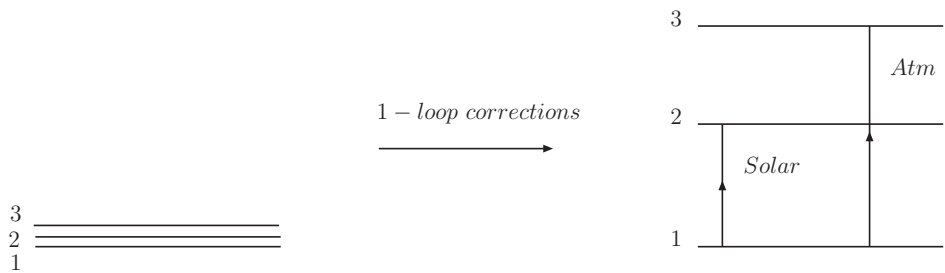} 
   \caption{Neutrino mass scales :  loop-level dominance}
   \label{figlevel2}
\end{figure}
\end{itemize}
In either case, the solar mass$^2$ difference originates from loop
effects from the lepton number violating parameters in (\ref{lnvp}).

The correct neutrino mass hierarchy can be always generated by the proper
choice of just two of the lepton number violating parameters from the
list of (\ref{lnvp}) -- one of which sets the scale of the atmospheric
mass$^2$ difference, the second setting the solar mass$^2$ difference.
Of course, in the most general case all parameters can contribute.

After choosing the lepton number violating (LNV) parameters, the method described in
section~\ref{subsec:inputpar} is then employed to determine the
charged lepton Yukawa matrix.  In general it needs to be non-diagonal,
in order to reproduce correct masses of both neutral and charged
leptons and the $U_{MNS}$ mixing matrix. The non-diagonal Yukawa
matrix (thus also non-diagonal charged lepton mass matrix), may easily
give rise to effects which are already subject to strong experimental
bounds; tree level lepton flavour processes, such as $\mu \to e\gamma$
or $\mu \to eee$, are not suppressed and loop corrections to the
electron decays will have contributions proportional to the tau mass.
To avoid such problems, the specific cases considered in
the next sections are those for which the large mixing in the lepton
sector, as seen in the MNS matrix, has its origin purely in the
neutral sector, and the charged lepton Yukawa couplings remain
flavour-diagonal.  The formalism we have described thus far allows the
correct masses and mixing of charged leptons to be initialised.  However, this
will lead, in general, to an off-diagonal lepton Yukawa matrix.  These, less
natural, initial parameters are not necessarily ruled out and within the framework
set out above, it is entirely possible to perform the calculations as
described.  However, we now prefer to consider a set of parameters for which we
do not rely on cancellations in the charged lepton sector to make the model
phenomenologically viable.  The simplest way in which this can be achieved, is
to find LNV parameters for which lepton Yukawas are diagonal.

From Eq.~(\ref{eq:mns}), for the case where the lepton Yukawa is
diagonal and therefore ${\cal Z}_l$ is the unit matrix, we see that
\begin{eqnarray}
	{\cal{Z}}_{\nu} = U_{{MNS}}^{\dagger} \;,
\end{eqnarray} 
up to higher order terms.  Using the MNS matrix as an input, it is
possible to see which ratios of entries in the mass matrix give rise
to the correct leptonic mixing, being,
\begin{eqnarray}
	m_\nu^{\textup{\tiny{eff}}}= {\cal{Z}}_{\nu}^{*}\:
	\textup{diag}(m_1,m_2,m_3) \: {\cal{Z}}_{\nu}^{\dagger}
= m_k \: U_{{MNS}\,ki} \: U_{{MNS}\,kj}\;.
\end{eqnarray}
For example, to set the atmospheric scale at tree level, we can see
from Eq.~(\ref{eq:eff}) that as long as the hierarchy of $\mu_i$
matches the ratios of any of the rows in the MNS matrix, the mass
matrix follow the correct pattern to be consistent with the observed
mixing matrix.  If a second mass scale is set up using the pattern
dictated by one of the other rows of the MNS matrix, the full MNS
matrix will be produced upon diagonalisation.

It is worth noting the nontrivial fact that such an approach,
i.e. generating correct structure of neutrino masses and mixings,
while keeping FCNC processes in charged lepton sector suppressed, is
at all possible.


\subsection{Tree level dominance scenario}

\begin{figure}[t] 
   \centering
   \includegraphics[width=6in,bb=50 50 410 302]{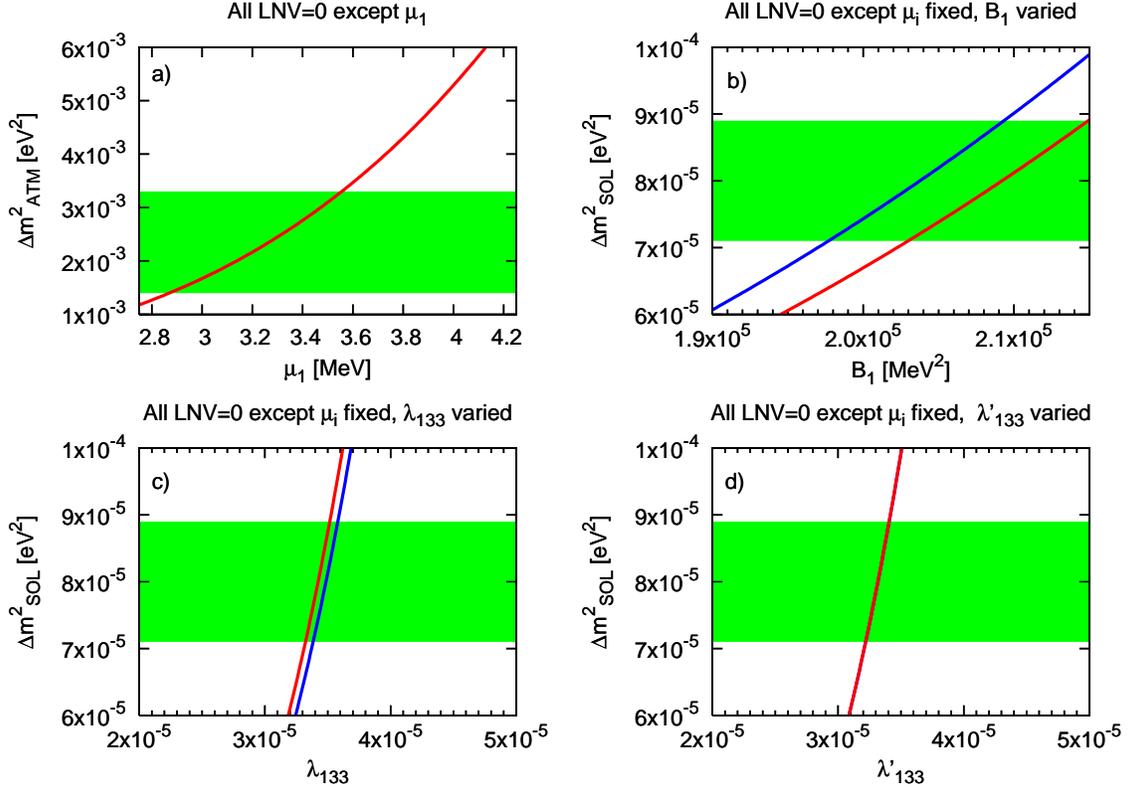} 
   \caption{Predictions for atmospheric and solar neutrino mass$^2$
   differences ($\Delta m^2_{\rm ATM}$) and ($\Delta m^2_{\rm SOL}$)
   for the {\it tree level dominance} scenario vs. variations of
   Lepton Number Violating (LNV) parameters as displayed in figure
   titles. The 3$\sigma$ gray (green) band consistent with experiment
   is displayed for comparison, as well as the full (gray or red
   curve) and approximate (dark or blue curve) results.  a) Only
   $\mu_1$ is varied. For all other figures, $\mu_i$ is fixed as in
   hierarchy (A) as explained in the text and b) $B_1$, or c)
   $\lambda_{133}$, or d) $\lambda'_{133}$, is varied respectively.  }
   \label{fig01}
\end{figure}

At tree level, the mass of a neutrino can be set using $\mu_i$
parameters (\ref{nmass}).  The top left panel in Fig.~\ref{fig01}
shows how the value of ($\Delta m^2_{\rm ATM}$) varies with $\mu_1$
only, setting $\mu_{2,3}=0$.  The grey (or red in color) line is the
result given by diagonalising the full neutralino matrix in
(\ref{eq:mnsplit}) and the dark (blue in color) line is given by
(\ref{nmass}). They agree perfectly in Fig.~\ref{fig01}a and thus only
one line is shown.  The shaded band shows the current 3$\sigma$
limits.  From Eq.~(\ref{nmass}) it can be seen, however, that it is
$|\mu_1|^2+|\mu_2|^2+|\mu_3|^2$ which sets the mass of the tree level
neutrino and as such it is straightforward to set any hierarchy
between the $\mu_i$ and still maintain the same value for the
atmospheric difference.
To correctly reproduce the MNS matrix, we choose as an input a 
very simple hierarchy between the $\mu_i$ parameters,
\begin{eqnarray}
{\rm Hierarchy ~(A) ~:}\qquad
\mu_1=\frac{\mu_2}{\sqrt{2}}=\frac{\mu_3}{\sqrt{3}} \;.
\label{mus}
\end{eqnarray}
The scale of all three $\mu_i$ is set such that a tree level neutrino
of the correct mass is generated which result in the observed
atmospheric mass difference, being,
\begin{eqnarray}
\mu_1=1.47~\textup{MeV}\;, \qquad \mu_2=\sqrt{2}\times1.47~\textup{MeV}\;,
\qquad \mu_3=\sqrt{3}\times1.47~\textup{MeV} \;.
\label{musval}
\end{eqnarray}
At tree level, this choice of hierarchy gives rise to the MNS matrix,
up to the $SU(2)$ rotation described earlier, being driven solely by
the neutral sector; the charged lepton mass matrix is diagonal, and as
such we have chosen a set of parameters within this basis which avoids
the possible phenomenological problems.

A further, single lepton number violating parameter can then be chosen
to set the scale of the solar mass$^2$ difference.  The question of
the arbitrariness of the tree level neutrino basis is complicated by
the requirement that once the loop corrected mass matrix is
diagonalised, ${\cal Z}_l$ being the unit matrix is consistent with
the experimentally observed MNS matrix.  As only one further lepton
number violating coupling is initiated, the ratios in which the loop
effects are distributed in the loop corrected mass matrix are
approximately determined by the tree level mixing matrix.  As such we
can determine an approximate expression for the extra contribution to
the full rotation matrix from rediagonalising the loop effects.  The
further condition that the full rotation must reproduce the MNS matrix
allows us to fix the tree level basis.

The three further Figs.~\ref{fig01}(b,c,d) show the range of possible
parameters in this scenario.  In each of these plots, the set of
$\mu_i$ are given the values (\ref{musval}) and another, single lepton
number violating coupling is varied.  In each case, the gray (or red)
line shows the full result and the dark (blue) line is the result
predicted by the approximate solutions.  The fact that $\lambda_{133}$ and
$\lambda'_{133}$ give the correct solar mass difference over a similar range
of parameters is a numerical coincidence.  For this example, the factors from the
different fermion masses in the quark loop, colour counting and scalar mixing
cancel each other.

The contribution of the neutral scalar loop discussed in
section~\ref{sec:3}, results from cancellations between the CP-even
and the CP-odd diagrams and may includes contributions of
approximately the same order.  As such, the approximation presented
earlier in the text, Eq.~(\ref{fneutral2}), does not agree well with
the full result.  The discrepancy between the full result and the
approximate result reflects the fact that various contributions arise
from different places in the full calculation (e.g. the effect on the
mixing matrices, the effect on the sneutrino masses).  The approximate
result plotted in Fig.~\ref{fig01}b is given by
\begin{eqnarray}
\Sigma^D_{N\, pp}&\simeq& -\sum_{r=1}^{7} \sum_{i,j=1}^{3}
 \frac{m_{\Kz{r}{}}{\cal Z}_{\nu \, ip}^2}{4(4\pi)^2} \left [
 \frac{e}{c_W} {\cal Z}_{N\, 1r} - \frac{e}{s_W} {\cal Z}_{N\, 2r}
 \right ]^2
%
 \left( \left[ \frac{\Delta m_{\tilde{\nu} i}^2 }{(m_{\tilde{\nu} i}^2
 - m_{\Kz{r}{}}^2 )} -\frac{ m_{\Kz{r}{}}^2 \Delta m_{\tilde{\nu} i}^2
 }{(m_{\tilde{\nu} i}^2 - m_{\Kz{r}{}}^2 )^2}
 \ln\frac{m_{\Kz{r}{}}^2}{m_{\tilde{\nu} i}^2} \right] \delta_{ij}
 \right. \nonumber\\
&+& \left[ \frac{B_i^2 \sin^2(\beta-\alpha)}{\cos^2 \beta
 (M_H^2-M_i^2)^2} + \frac{B_i^2 \cos^2(\beta-\alpha)}{\cos^2 \beta
 (M_h^2-M_i^2)^2} - \frac{B_i^2 \cos^2\beta}{(M_A^2-M_i^2)^2} \right]
 \delta_{ij} \frac{M_i^2 \ln M_i^2 - m_{\Kz{r}{}}^2 \ln
 m_{\Kz{r}{}}^2}{M_i^2-m_{\Kz{r}{}}^2} \nonumber \\
&+& \left[ \frac{B_i B_j \cos^2(\beta-\alpha) }{\cos^2\beta
 (M_i^2-M_h^2)(M_j^2-M_h^2)} \right] \frac{M_h^2 \ln M_h^2 -
 m_{\Kz{r}{}}^2 \ln m_{\Kz{r}{}}^2}{M_h^2-m_{\Kz{r}{}}^2} \nonumber \\
&+& \left[ \frac{B_i B_j \sin(\beta-\alpha) }{\cos^2\beta
(M_i^2-M_H^2)(M_j^2-M_H^2)} \right] \frac{M_H^2 \ln M_H^2 -
m_{\Kz{r}{}}^2 \ln m_{\Kz{r}{}}^2}{M_H^2-m_{\Kz{r}{}}^2} \nonumber \\
&-& \left.\left[ \frac{B_i B_j }{\cos^2\beta (M_i^2-M_A^2)(M_j^2-M_A^2)}
  \right] \frac{M_A^2 \ln M_A^2 - m_{\Kz{r}{}}^2 \ln
  m_{\Kz{r}{}}^2}{M_A^2-m_{\Kz{r}{}}^2} \right)\;. 
\label{fneutralnew}
\end{eqnarray}
The approximate result for the charged scalar loop, given by
Eq.~(\ref{alamcon}) agrees well with the full result
(Fig.~\ref{fig01}c).  However, as $\lambda_{133}=-\lambda_{313}$,
there are other diagrams which contribute to the full result which are
not included in the approximation.  The approximate expression
captures the important effect.  The agreement between the full result
and the approximate result, given by Eq.~(\ref{aqsq}) when varying
$\lambda'_{133}$ (see Fig~\ref{fig01}d) is very good, as the diagram
highlighted in the text is the only diagram which contributes.

\subsection{Loop level dominance scenario}

\begin{figure}[t] 
   \centering
   \includegraphics[width=6in,bb=50 50 410 302]{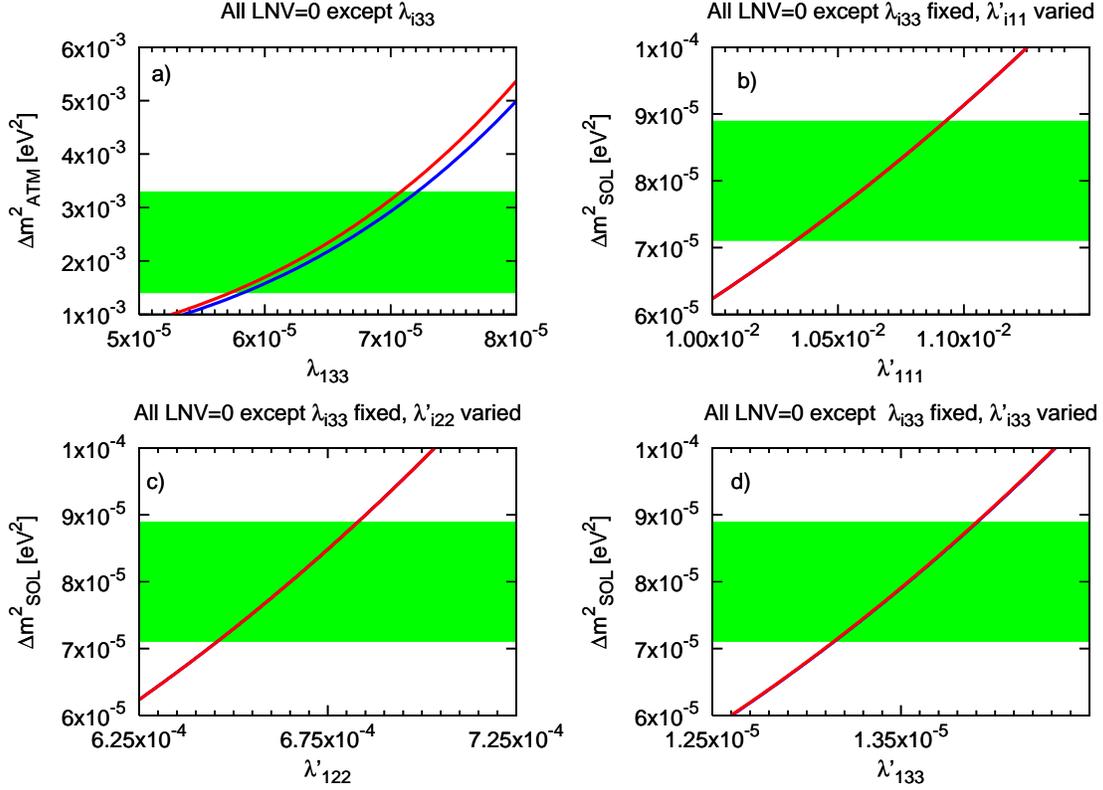} 
   \caption{Same as in Fig.~3 but for the {\it loop-level dominance}
    scenario. All LNV parameters are zero apart from a)
    $\lambda_{i33}$ that is varied in hierarchy (D). For all other
    figures, $\lambda_{i33}$ is fixed to a value (see the text)
    consistent with the atmospheric mass$^2$ difference and b) only
    $\lambda'_{i11}$ is varied in hierarchy (B) or c) only
    $\lambda'_{i22}$ in hierarchy (B) or d) only $\lambda'_{i33}$ in
    hierarchy (B) in order to accommodate the solar mass$^2$
    difference. }
   \label{fig03}
\end{figure}

It is possible for both the solar and the atmospheric scales to be set
by loop corrections.  This happens if the bilinear parameters $\mu_i$
are small enough. In this section we analyse this case setting
strictly $\mu_1=\mu_2=\mu_3=0$, so that the one-loop corrections to
the full $3\times 3$ neutrino mass matrix are finite. Otherwise a more
involved renormalisation scheme has to be implemented.

Again, we would like to set the Lagrangian parameters such that one can
generate the correct structure of the MNS matrix while keeping the charged
Yukawa couplings flavour-diagonal.  This can be achieved if the neutrino
mass hierarchy is governed by the trilinear $\lambda$ and $\lambda'$
couplings.  For the diagrams dominated by trilinear couplings the
flavour of the external legs of the loop can be ``swapped
independently'' of the flavour of the particles in the loop, just
changing the appropriate indices of the $\lambda,\lambda'$ matrices in the
loop vertices.  Setting the $\lambda$ and $\lambda'$ entries which
control the couplings of the external legs in certain hierarchies, one
can ensure that also the ratios of the various entries in the one loop
corrected neutrino mass matrix are such that they give rise to the
correct $U_{MNS}$ rotation matrix.

\begin{figure}[t] 
   \centering
   \includegraphics[width=6in,bb=50 50 410 302]{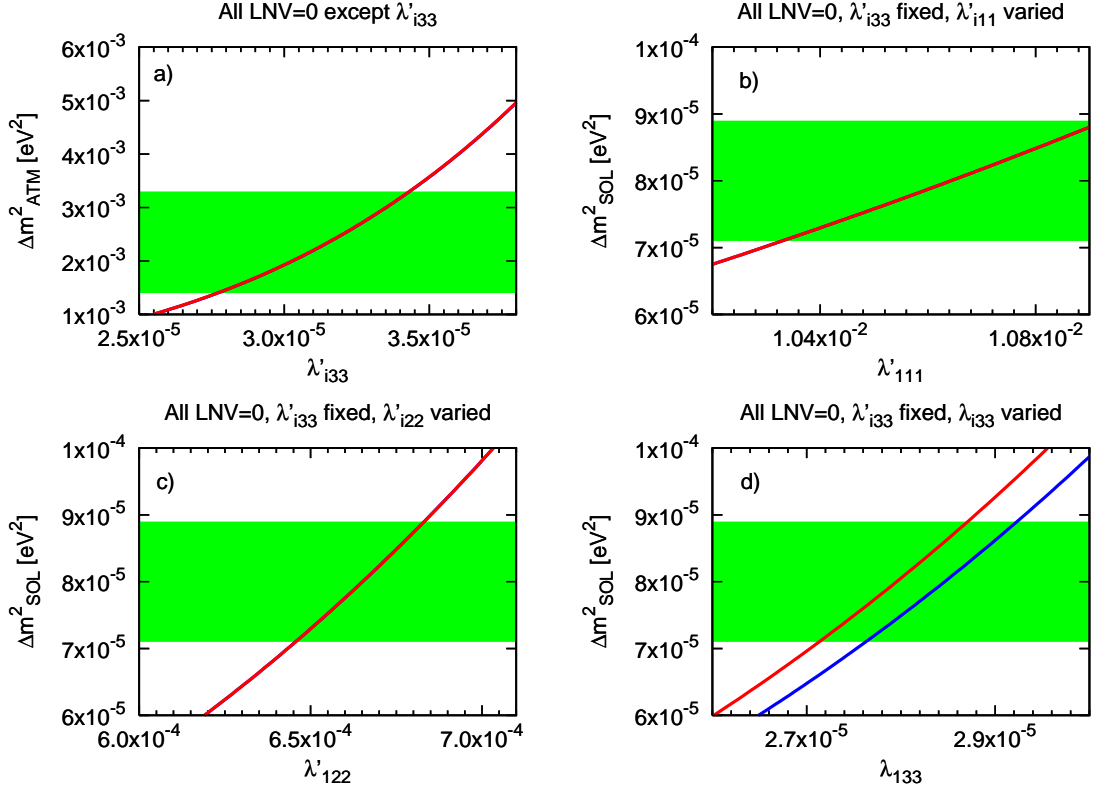} 
   \caption{Same as in Fig.~3 but for the {\it loop-level dominance}
    scenario. All LNV parameters are zero apart from a)
    $\lambda'_{i33}$ that is varied in hierarchy (B). For all other
    figures, $\lambda'_{i33}$ is fixed to a value (see the text)
    consistent with the atmospheric mass$^2$ difference and b) only
    $\lambda'_{i11}$ is varied in hierarchy (C) or c) only
    $\lambda'_{i22}$ in hierarchy (C) or d) only $\lambda_{i33}$ in
    hierarchy (D) in order to accommodate the solar mass$^2$
    difference.}
   \label{fig04}
\end{figure}

The possible hierarchies are given by the rows of the MNS matrix and
are, with a generic coupling $\lambda^{(')}_{ijj}$ as
follows\footnote{Couplings with $\lambda^{(')}_{ijk}$, with $j\ne k$
have only negligible contributions to neutrino masses and excluded
from our hierarchy list.},
\begin{eqnarray}
{\rm Hierarchy ~(B) ~:}& \qquad        \lambda'_{1jj}=\frac{\lambda'_{2jj}}{\sqrt{2}}=\frac{\lambda'_{3jj}}{\sqrt{3}} \\
{\rm Hierarchy ~(C) ~:}& \qquad       \lambda'_{1jj}=\frac{\lambda'_{2jj}}{\sqrt{2}}=-\frac{\lambda'_{3jj}}{\sqrt{3}} \\
{\rm Hierarchy ~(D) ~:}& \qquad	\lambda_{1jj}=-\sqrt{2} \lambda_{2jj}\quad 
\;, \quad \lambda_{3jj}=0  \;.
\label{hierarchy}
\end{eqnarray}
Due to the antisymmetry of the first two indices of $\lambda$ it can
only be chosen to follow hierarchy (D) described above.

As the nature of the loop corrections due to $B_i$ means that the
external legs cannot be swapped without affecting the flavour
structure inside the loop, it is difficult to fix a hierarchy of $B_i$
in the Lagrangian which will automatically give rise to the correct
ratios in the one loop corrected mass matrix.

We consider first, the case where the atmospheric mass$^2$ difference
is set by $\lambda_{i33}$ in hierarchy (D).  The range of values for
which the correct atmospheric mass difference is given is plotted in
Fig.~\ref{fig03}a.  Note that although we plot on the x-axis
$\lambda_{133}$, the coupling $\lambda_{233}$ is also varying to keep
the hierarchy (D) fixed.  The fact that both $\lambda_{133}$ and 
$\lambda_{233}$ contribute is the reason the value of the coupling is only a
little greater than the value of $\lambda_{133}$ which correctly reproduces
the solar mass difference in the tree-level dominated scenario.

The further three panels [Fig.~\ref{fig03}(b-d)] have a fixed set of
$\lambda_{i33}$ in hierarchy (D) giving the atmospheric difference.
Being,
\begin{eqnarray}
	\lambda_{133}=6.7\times10^{-5} \;,
	\qquad \lambda_{233}=-\frac{6.7\times10^{-5}}{\sqrt{2}}\;,
	\qquad	\lambda_{333}=0 \;.
\label{lamsval}
\end{eqnarray}
$\lambda_{333}=0$ due to the antisymmetry between the first two
indices, fitting hierarchy (D).  In addition to this,
Fig.~\ref{fig03}b varies $\lambda'_{111,211,311}$ in hierarchy (B)
giving the solar mass$^2$ difference.  Again, we plot the solar
mass$^2$ difference against the value of $\lambda'_{111}$, however
$\lambda'_{211,311}$ are being varied at the same time.  The two
remaining panels take in turn different sets of three $\lambda'$
couplings, $\lambda'_{i22}$ and $\lambda'_{i33}$.  The final two
indices determine which particle is produced in the loop.  As such,
with a lighter particle in the loop, the couplings must be greater to
compensate.  We see that, in moving from one panel to the next
Fig.~\ref{fig03}(b$\to$ d), to produce the same mass difference, a
smaller value of the coupling is required with a heavier particle in
the loop.  With the down quark in the loop (Fig.~\ref{fig03}b) the
value needed for the coupling may result in large contributions to the
neutrinoless double beta decay rate as it is already approaching the
excluded regime~\cite{ADD99}.

Finally, Fig.~\ref{fig04}a, we show how the atmospheric mass$^2$
difference can be set by the three $\lambda'_{i33}$ couplings in
hierarchy (B), plotting the result for the atmospheric mass$^2$
difference against $\lambda'_{133}$.
Next, we set $\lambda'_{i33}$ to take the following values 
\begin{eqnarray}
	\lambda'_{133}=3.25\times10^{-5}\;,
	 \qquad \lambda'_{233}=\sqrt{2}\times 3.25\times10^{-5}\;,	\qquad	\lambda'_{333}=\sqrt{3}\times 3.25\times10^{-5} \;.\nonumber \\
\label{lampsval}
\end{eqnarray}
The remaining three plots, Fig.~\ref{fig04}(b,c,d), show the change in
solar mass$^2$ difference, as sets of either $\lambda'_{i11,i22}$ in
hierarchy (C) or the set $\lambda_{i33}$ in hierarchy (D) are varied.

\section{Conclusions}
\label{sec:5}

An increasingly accurate picture of the neutrino sector, with masses
much smaller than the charged leptons and a distinctive mixing matrix
in the $W$-vertex, is being discerned by current experiments.  We note
that there are three preferred ${\cal Z}_N$ symmetries in the
supersymmetric extension of the Standard Model with minimal particle
content.  Imposing a ${\cal Z}_2$ symmetry results in the widely
studied R-parity conserving MSSM, however another preferred symmetry,
${\cal Z}_3$, gives rise to a Lagrangian which explicitly violates
lepton number.  These interactions lead to neutrino masses, both
through a `see-saw' type suppression at tree level and through
radiative corrections.  We have considered the most general scenario
in this model; no assumptions have been made concerning CP-violation
or intergenerational mixing, for example.

We present, in Appendix~\ref{app:mass}, the tree level mass matrices
of the model and, in Appendix~\ref{app:feyn}, the full set of Feynman
rules for the neutral fermion interactions.  The calculation has been
performed using two-component Weyl fermion notation;
Appendix~\ref{app:weyl} contains a derivation of the propagators in
this notation, included primarily for pedagogical reason, and generic
self-energy diagrams for scalar and gauge boson corrections to
fermions.

In the basis set out in our previous work~\cite{DRRS}, we find that a
non-zero neutrino mass will arise at tree level unless all $\mu_i$ are
zero and analyse in detail, the further contributions to masses that
come from loop corrections.  We show that the magnitude of the
contributions due to neutral fermion loops, examined in
section~\ref{subsubsec:neutral} are determined by the size of the
bilinear supersymmetry breaking parameter, $B_i$; that loops with
charged fermions, described in section~\ref{subsubsec:charged}, have a
contribution due to trilinear lepton number violating couplings in the
superpotential, $\lambda_{ijk}$; and that quark loops,
section~\ref{subsubsec:quark}, are determined by the trilinear lepton
number violating coupling $\lambda'_{ijk}$.  Each of these
contributions can be dominant.  In sections~\ref{subsubsec:z}
and~\ref{subsubsec:w}, we consider the gauge loops and why they do not
give large contributions to neutrinos which are massless at tree
level.  We derive expressions for the full calculation, which form the
basis of our numerical analysis.  We also present approximate
expressions in each section, which are simple, compact formulas
encoding the important information pertaining to each diagram.  In our
presentation of the results, as seen in
Figs.~\ref{fig01},\ref{fig03},\ref{fig04} these simple expressions are
shown to be in good agreement with the full result.

The lepton sector in the \LMSSM is much more involved than the lepton
number conserving MSSM.  Mixing between leptons, gauginos and
higgsinos ensures the question of initialising Lagrangian parameters
must be carefully considered.  A framework has been constructed in
section~\ref{sec:2} to correctly reproduce the charged lepton masses
and MNS matrix for any set of lepton number violating couplings.

In constructing the framework in which to perform the calculation it
is clear that there will be, in general, large intergenerational
mixing in the lepton Yukawa matrix, this allows the possibility of
unsuppressed tree level flavour violating processes, already subjected to strong
bounds.  To circumvent this problem we considered sets of Lagrangian
parameters for which the MNS matrix has its origin solely in the
neutral sector, the lepton Yukawa matrix being diagonal.  The three
rows of the MNS matrix correspond to three sets of ratios between
entries in the loop corrected mass matrix which will give the correct
MNS angles.  It is possible to set these ratios by setting hierarchies
in the couplings between generations.  With the condition that it must
be possible to change the flavour of the external legs of the diagram
without affecting the flavour structure of the loop, there is some
freedom in choosing which group of Lagrangian parameters we set in
each hierarchy.

Lepton number conserving parameters were fixed to be the SPS1a
benchmark point, and we have investigated the effect of varying the
lepton number violating couplings, as seen in
Figs.~\ref{fig01},\ref{fig03},\ref{fig04}.  We have shown that values
for lepton number violating couplings exist, which give the correct
atmospheric and solar mass$^2$ difference, charged lepton masses and
mixing, which are not already excluded by existing studies of low energy bounds.  There are
two distinct scenarios that achieve this: the tree level dominance
scenario, in which the atmospheric scale is set at tree level and the
solar scale set by radiative effects, and another, the loop level
dominance scenario, in which both the atmospheric and solar scales are
set by radiative corrections.

In the tree level dominance scenario, we choose the $\mu_i$ parameters
to be of the order of 1 MeV,
such that the correct result for the atmospheric mass$^2$ difference
is obtained.  They are chosen to obey a certain hierarchy, which
ensures the mixing matrix is consistent with observed MNS.

In addition to this, a further, single lepton number violating
coupling can set the scale of the solar mass$^2$ difference by
determining the contribution of the appropriate loop diagram.  It is
possible to generate loop diagrams of the appropriate scale, by
including either a non-zero $\lambda$,$\lambda'$ or $B$ coupling.  We
find that the correct solar scale can then be set by any of
\begin{eqnarray}
	B_1 &\sim& 0.21 \, \textup{GeV}^2 \sim \Big[\,300\,
	\mu_{1}\,
	\Big]^2 \; , \nonumber \\
	\lambda_{133} &\sim& 3.4 \times 10^{-5} \sim y_e \; , \nonumber \\
	\lambda'_{133} &\sim& 3.2 \times 10^{-5} \sim 0.1  y_d \; , 
\end{eqnarray}
where $y_{e,d}$ is the Yukawa coupling of either the electron or the down
quark, presented here merely for the sake of comparison.

In the second case, the correct masses and mixing for both charged and
neutral fermions can be achieved without a massive neutrino at tree
level.  The solar and atmospheric mass$^2$ differences both arise from
radiative corrections at one loop, using loop contributions whose
value is determined by sets of $\lambda$ or $\lambda'$ couplings in
given hierarchies, such that the observed MNS is generated.  Firstly,
we show that we can set the atmospheric scale with a set of $\lambda$
couplings of the order of the electron Yukawa coupling,
then find the solar scale is correctly set by $\lambda'$ couplings of
the order of the down quark Yukawa coupling.

Alternatively, the atmospheric scale can be set by $\lambda'$
couplings,
\begin{eqnarray}
	\lambda'_{133}=\frac{\lambda'_{233}}{\sqrt{2}}=\frac{\lambda'_{333}}{\sqrt{3}}= 3.25\times10^{-5} \sim 0.1  y_d  \;,
\end{eqnarray}
and the solar mass$^2$ difference can be generated by another set of
$\lambda'$ couplings,
\begin{eqnarray}
	\lambda'_{122}=\frac{\lambda'_{222}}{\sqrt{2}}=-\frac{\lambda'_{322}}{\sqrt{3}}
\sim 6.5\times10^{-4} \sim 2 y_d \;,
\end{eqnarray}
or a set of $\lambda$ couplings of the order of the electron Yukawa.

We include some comments on how this work compares with previous work
in the literature.  We highlight where our results agree with
statements made in the literature and comment on results presented in
different bases.

The study of neutrino masses will provide the basis for further work
concerning lepton number violating phenomena.  The ranges of values
for lepton number violating parameters required to produce the correct
masses and mixing, will be reflected in processes such as tree level
lepton flavour violating decays and will have repercussions concerning
rare events such as neutrinoless double beta decay.  This will make a
valuable link between collider experiments and upcoming neutrino
experiments.  In this paper, we have made a framework for these
investigations.

\acknowledgments{AD would like to thank the Nuffield Foundation for
financial support.  SR acknowledges the award of a PPARC studentship.
We would like to thank Max Schmidt-Sommerfeld for collaboration at the
early stages of this work.  AD and SR would like to thank W.  Porod
for discussions on~\cite{ValleD68} and H.  Haber and H.  Dreiner for
discussions during the Pre-SUSY'05 Meeting relevant to
\cite{preSUSY}. The work of JR was supported in part by the Polish
Committee of Scientific Research under the grant number 1~P03B~108~30
(2006-2008). JR would also like to thank to IPPP Durham for the
hospitality during his stay there.}

\newpage


\appendix
\renewcommand{\theequation}{\Alph{section}-\arabic{equation}}

\noindent {\Large \bf Appendix}

\section{The  Lagrangian and the mass matrices of the \LMSSM} 
\label{app:mass}
\setcounter{equation}{0} 
 
We strictly follow the notation of Refs.~\cite{Allanach,DRRS}.  The
\LMSSM superpotential is given by Eq.~(\ref{superpot1}).  The main
discussion of this paper is confined in the lepton sector, but for
completeness we define here also the mass basis in the quark sector.
To this end, we rotate the four quark superfields to a basis where
both $\lamp{0ij}$ and $\YU{ij}$ are diagonal
\begin{eqnarray}
D_L' \longrightarrow \ZdLs{} D_L\;, & \qquad & \bar{D}' \longrightarrow \ZdR{} \bar{D} \;, 
\nonumber  \\[2mm]
U_L' \longrightarrow \ZuL{} U_L \;, & \qquad & \bar{U}' \longrightarrow \ZuRs{} \bar{U} \;.
\label{urot}
\end{eqnarray}
By absorbing a rotation matrix into the Lagrangian parameters, one can
write down the superpotential (\ref{superpot1}) as
\begin{eqnarray}
 \SPot &=& 
\half \ep{ab} \lam{\alpha \beta j} \Lx{\alpha}{a} \Lx{\beta}{b} \bar{E}_{j} 
+  \lamp{\alpha ij}  \Lx{\alpha}{1} \Q{i}{2}{x} \bar{D}_{j}^{x} 
-  \lamp{\alpha kj} K^*_{ik}  \Lx{\alpha}{2} \Q{i}{1}{x} \bar{D}_{j}^{x} 
- \ep{ab} \mux{\alpha} \Lx{\alpha}{a} \HB{}{b} \nonumber \\[2mm]
&+&\YU{ij}  \Q{i}{1}{x} \HB{}{2} \bar{U}_{j}^{x} 
-\YU{kj} K_{ki}  \Q{i}{2}{x} \HB{}{1} \bar{U}_{j}^{x} \;, \label{superpot2}
\end{eqnarray} 
where $\lamp{0jk}$ and $\YU{ij}$ are diagonal matrices and ${\cal
L}^1$ (${\cal L}^2$) is the neutrino (electron) component of the
$SU(2)$ doublet.  The charged current part of the Lagrangian,
\begin{equation}
 {\cal L}_W = \frac{e}{\sqrt{2} s_W} \bar{u}_{L i} K_{ij}
 \bar{\sigma}^\mu W_\mu^+ d_{Lj} + {\rm H.c}\;, \end{equation}
 diagrammatically reads (in Weyl spinor notation) as,
\begin{eqnarray}
\ffv{\uL{i}{}{}}{\dL{j}{}{}}{\W{\mu}{}}
{:\,  \frac{ie}{\sqrt{2} s_W }  K_{ij}  \sigb{}{\mu}  }{}{}{}{}{}{}{} \label{Wvertex}
\end{eqnarray}
with $K_{ij} = \ZuLs{ki} \ZdLs{kj}$ being the CKM matrix.  We rotate
all fields in the basis where sneutrino vevs are zero.  Then the soft
supersymmetry breaking terms are,
$$ 
\Lag_{\rm SSB} =
 - \left(M^2_{\tilde{\rm L}} \right)_{\alpha \beta} \snuLs{\alpha} \snuL{\beta} 
- \left(M^2_{\tilde{\rm L}}  \right)_{\alpha \beta} \seLs{\alpha} \seL{\beta}
- m_{\HB{}{}}^2 \shbzs \shbz
-  m_{\HB{}{}}^2 \shbps \shbp
$$
$$
-  \left(m_{\sEc{}{}}^2 \right)_{ij} \seRs{i} \seR{j}   
- \left( m_{\sQ{}{}}^2 \right)_{kl} K_{ik} K^*_{jl}   \suLs{i}{z} \suL{j}{z}
- \left( m_{\sQ{}{}}^2 \right)_{ij} \sdLs{i}{z} \sdL{j}{z}
$$
$$
- \left( m_{\sDc{}{}}^2 \right)_{ij} \sdRs{i}{z} \sdR{j}{z}
- \left( m_{\sUc{}{}}^2 \right)_{ij} \suRs{i}{z} \suR{j}{z}
$$
$$ \bigg[
- \left(h_u \right)_{ij} \suL{i}{y} \shbz \suRs{j}{y}
+ \left(h_u \right)_{kj} K_{ki} \sdL{i}{y} \shbp \suRs{j}{y}
- h_{\alpha \beta k} \snuL{\alpha} \seL{\beta} \seRs{k} 
- h'_{\alpha j k} \snuL{\alpha} \sdL{j}{} \sdRs{k}{} 
\bigg.
$$
$$ 
+ h'_{\alpha k j} K^*_{ik} \seL{\alpha} \suL{i}{} \sdRs{j}{}
+B_\alpha \snuL{\alpha} \shbz 
- B_\alpha \seL{\alpha} \shbp 
$$
\begin{equation} \bigg.
+ \half M_1 \Bino{}{}\Bino{}{} 
+ M_2 \Wino{}{+} \Wino{}{-} 
+ \half M_2 \Wino{}{0} \Wino{}{0}
+ \half M_3 \Gino{}{R} \Gino{}{R}
+ \textup{H.c.}   \bigg] \;, \label{soft}
\end{equation}
where $B_\alpha$ is the four-component bilinear term $B_\alpha = (B_0,
B_i)$ and $h,h'$ are trilinear soft breaking couplings.  In the basis
of \cite{DRRS}, in addition to vanishing sneutrino vevs, one obtains
also diagonal sneutrino soft breaking mass terms,
$\left(M^2_{\tilde{\rm L}} \right)_{ij} \equiv
\left(\hat{M}^2_{\tilde{\rm L}} \right)_{ij} $ is given in Eq.~(2.25)
of \cite{DRRS}.  Notice also that the soft breaking mass which
corresponds to the mixing of the Higgs with the slepton is just the
term $B_i \tan\beta$.  In this basis, we shall now present the
spectrum of the model.  The neutral Higgs sector and approximate
formulae has been displayed in Ref.~\cite{DRRS}; we repeat only the
mass matrices here for completeness and definition.

\subsection{Mass Terms for CP-even Neutral Scalars}

After electroweak symmetry breaking, sneutrinos,
$\{\tilde{\nu}_{L_i}\}$, mix with Higgs bosons, $\{\tilde{\nu}_{L 0} =
h_1^0, h_2\}$, resulting in CP-even and CP-odd scalars (recall that
CP-symmetry is preserved at tree level even in the most general
R-parity violating MSSM~\cite{DRRS}).  The corresponding CP-even
neutral scalar mass terms are
$$
{\cal L} \ \supset \ - \left( \begin{array}{ccc} \mathrm{Re}\,h_0^2 &
  \mathrm{Re}\,\tilde\nu_{L0} &  \mathrm{Re}\,\tilde\nu_{Li} \end{array} \right)
  \ZRs{}  \ZRT{} \mathcal{M}^2_{\rm H} \ZR{} \ZRd{}
  \left( \begin{array}{c} \mathrm{Re}\,h_0^2
  \\
\mathrm{Re}\,\tilde\nu_{L0} \\
\mathrm{Re}\,\tilde\nu_{Lj}
\end{array} \right) \;,
$$
where 
\begin{eqnarray}
 \mathcal{M}^2_{H} = 
\left( \begin{array}{ccc}
{ \cos^2\beta M_A^2 + \sin^2 \beta M_Z^2 } &
{ - \half \sin 2 \beta ( M_A^2 +  M_Z^2 ) } &
{ - B_j } 
\\[3mm]
{ - \half \sin 2 \beta ( M_A^2 + M_Z^2 ) } &
{ \sin^2 \beta M_A^2 + \cos^2 \beta M_Z^2 } &
{  B_j \tan\beta}
\\[3mm]
{ - B_i } &
{  B_i \tan\beta} &
{ M^2_i \delta_{ij} }
\end{array} \right) \;,  \label{even}
\end{eqnarray}
%
and
\begin{equation}
M^2_i \  \equiv  \ (\hat{M}^2_{\tilde{\rm L}})_{i} + 
\half \cos2\beta M_Z^2 \;, \qquad M_A^2 = \frac{2 B_0}{\sin 2\beta} 
\;.  \label{mi}
\end{equation}
The rotation matrix is then given by
\begin{eqnarray}
\ZR{}^T \mathcal{M}^2_{\rm H}\ZR{} = {\rm diag}[m_{h^0}^2, m_{H^0}^2,
(m^2_{\tilde{\nu}_+})_i] \;.  \qquad i=1,...3 \;.\label{zrdef}
\end{eqnarray}
An approximate formula for the matrix $\ZR{}$ is given in Eq.~(3.9) of
Ref.~\cite{DRRS}.

\subsection{Mass Terms for CP-odd Neutral Scalars}

Mass Terms for CP-odd Neutral Scalars can be read from
$$
{\cal L} \ \supset \ - \left( \begin{array}{ccc} \mathrm{Im}\,h_0^2 &
  \mathrm{Im}\,\tilde\nu_{L0} &  \mathrm{Im}\,\tilde\nu_{Li} \end{array} \right)
  \ZAs{} \ZAT{}  \mathcal{M}^2_{\rm A} \ZA{} \ZAd{} 
  \left( \begin{array}{c} \mathrm{Im}\,h_0^2 \, 
\\ 
\mathrm{Im}\,\tilde\nu_{L0} \\
\mathrm{Im}\,\tilde\nu_{Lj}
\end{array} \right) \;,
$$
where
\begin{eqnarray}
 \mathcal{M}^2_{\rm A} = 
\left( \begin{array}{ccc}
{ \cos^2 \beta M_A^2 + \xi \sin^2 \beta M_Z^2 } &
{ \half \sin 2 \beta ( M_A^2 - \xi M_Z^2 ) } &
{ B_j } 
\\[3mm]
{ \half \sin 2 \beta ( M_A^2 - \xi M_Z^2 ) } &
{ \sin^2 \beta M_A^2 + \xi \cos^2 \beta M_Z^2 } &
{  B_j \tan \beta}
\\[3mm]
{ B_i } &
{ B_i \tan \beta} &
{ M^2_{i} \delta_{ij} }
\end{array} \right) \;, \label{matA}
\end{eqnarray}
and $\xi$ is the gauge-fixing parameter.  The rotation matrix $\ZA{}$
is defined through,
\begin{equation}
\ZA{}^T \mathcal{M}^2_{\rm A} \ZA{} = {\rm diag}[m_{G^0}^2, m_{A^0}^2,
(m_{\tilde{\nu}_-}^2)_i] \;, \qquad i=1,...3 \;.
\label{defza}
\end{equation}
An approximate formula for the matrix $\ZA{}$ is given in Eq.~(3.22)
of Ref.~\cite{DRRS}.

\subsection{Mass Terms for Charged Scalars}

In \LMSSM charged Higgs and charged sleptons mix.  The mass terms and
the rotation matrix $\ZH{}$, can be read from the Lagrangian
$$ \Lag \supset
- \left( \begin{array}{cccc} \shbps &\sham & \seL{j}& \seR{k} \end{array}\right)
	{\cal M}_{H^+}^2
\left( \begin{array}{c} \shbp \\ \shams\\ \seLs{i} \\ \seRs{l} \end{array}\right)=
- \left( \begin{array}{cccc} \shbps & \sham & \seL{j}& \seR{k} \end{array}\right)
	\ZH{} \ZHd{}
	{\cal M}_{H^+}^2
	\ZH{} \ZHd{}
	\left( \begin{array}{c} \shbp \\ \shams \\ \seLs{i} \\ \seRs{l} \end{array}\right)
$$
	\begin{equation}  
= - (\hat{{\cal M}}^2_{H^+})_{pq}  \Hps{p} \Hp{q} \;, \quad p,q=1,...8 \;, \label{defzhp}
\end{equation}
where the notation is self explanatory and
	$$
{\cal M}_{H^+}^2 = \left( \begin{array}{cc|} 
	M_A^2 \cos^2\beta
+M_W^2 \cos^2\beta
+\xi M_W^2 \sin^2\beta
&M_A^2 \sin\beta \cos\beta 
+ M_W^2 (1-\xi) \sin\beta \cos\beta
\\
M_A^2 \sin\beta \cos\beta 
+ M_W^2 (1-\xi) \sin\beta \cos\beta
&M_A^2 \sin^2\beta + \xi M_W^2 \cos^2\beta + M_W^2 \sin^2 \beta
\\
B_j
&B_j \tan\beta
\\
\frac{1}{\sqrt{2}} \lams{0ml} \mux{m} v_d
&\frac{1}{\sqrt{2}} \lams{0ml} \mux{m} v_u
\\
\end{array}\right.  
$$
\vspace{3pt}
\begin{equation}
\left.
\begin{array}{|cc}  B_i 
& \frac{1}{\sqrt{2}} \lam{0ml} \muxs{m} v_d \\
 B_i \tan\beta
&  \frac{1}{\sqrt{2}} \lam{0ml} \muxs{m} v_u \\
 M_i^2 \delta_{ij} - M_W^2 \cos^2 2\beta \: \delta_{ij} + \half \lams{0im} \lam{0jm} v_d^2
& \frac{1}{\sqrt{2}} \big( - \lam{\alpha jl} \muxs{\alpha} v_u + h_{0jl} v_d \big)   \\
\frac{1}{\sqrt{2}} \big( - \lams{\alpha jl} \mux{\alpha} v_u + h^*_{0jl} v_d \big)
&  \left(m_{\sEc{}{}}^2 \right)_{lk}    
+(M_W^2-M_Z^2) \cos^2 2\beta \: \delta_{lk}
+\half \lam{0ml}\lams{0mk} v_d^2  \\
\end{array} \right) \;.  \label{mat:ch}
\end{equation}

\subsection{Mass terms for down-type squarks}

Mass terms and rotation matrices for down-type squarks arise from the
Lagrangian part
 \begin{eqnarray} -{\cal L} \supset
\left( \begin{array}{cc} \sdLs{i}{z} & \sdRs{j}{z} \end{array} \right) 
	\Msq{d}{}
	\left( \begin{array}{c} \sdL{k}{z} \\ \sdR{l}{z}  \end{array} \right)
&=&  \left( \begin{array}{cc} \sdLs{i}{z} & \sdRs{j}{z} \end{array} \right) 
	\Zsds{} \ZsdT{}
	\Msq{d}{} 
	\Zsds{} \ZsdT{}
	\left( \begin{array}{c} \sdL{k}{z} \\ \sdR{l}{z}  \end{array} \right)
	\nonumber \\[2mm]
&=&  \Msqh{d}{pq} \: \sds{p}{z} \sd{q}{z}  \;,  \qquad p,q=1,...,6 \label{sdown}
	\end{eqnarray}
	where 
\begin{eqnarray}
\Msq{d}{} =
\left( \begin{array}{cc} 
 \left( m_{\sQ{}{}}^2 \right)_{ik} 
+\half \lamp{0 km}\lamps{0 im} \snuLvev{0}^2 
+(\frac{g^2}{24} + \frac{g_2^2}{8})(  \shbzsvev^2   
-\snuLsvev{0}^2 ) \delta_{ik} 
& -\isqt  \mux{\alpha} \lamps{\alpha il}\shbzvev  
+ \isqt h'^*_{0 i l} \snuLsvev{0} \\
- \isqt \lamp{\alpha kj} \muxs{\alpha} \shbzsvev  
+\isqt  h'_{0 kj} \snuLvev{0} 
&  \left( m_{\sDc{}{}}^2 \right)_{jl} 
+\half  \lamp{0 qj}\lamps{0 ql}\snuLvev{0}^2  
+\frac{g^2}{12}(\shbzsvev^2-\snuLsvev{0}^2) \delta_{jl}\\
\end{array} \right) \;.  \nonumber \\ \label{matd}
\end{eqnarray}
Recall that $\lamp{0 km}=\hat{Y}_{{D}k}\, \delta_{km}$ are diagonal
down-quark Yukawa couplings.

\subsection{Mass terms for up-type squarks}

The mass terms for up-type squarks are
\begin{eqnarray}
-{\cal L}= \left( \begin{array}{cc} \suLs{i}{z} & \suRs{j}{z} \end{array} \right)
	\Msq{u}{}
	\left( \begin{array}{c} \suL{k}{z} \\ \suR{l}{z}  \end{array} \right)
 & = &
 \left( \begin{array}{cc} \suLs{i}{z} & \suRs{j}{z} \end{array} \right)
	\Zsu{} \Zsud{}
	\Msq{u}{}
	\Zsu{} \Zsud{}
	\left( \begin{array}{c} \suL{k}{z} \\ \suR{l}{z}  \end{array} \right)
\nonumber \\[2mm]
  &=& 
  \Msq{u}{pq} \:  \sus{p}{z} \su{q}{z}\;, \qquad p,q=1,...,6 \;.
	\end{eqnarray}
where 
\begin{eqnarray}
&&\Msq{u}{} = \nonumber\\ 
&&\left( \begin{array}{cc} 
	\left( K m_{\sQ{}{}}^2 K^\dagger\right)_{ik}  
+\half  (Y_U Y_U^\dagger)_{ki} \shbzvev^2
+(\frac{g^2}{24} -\frac{g_2^2}{8}) (\shbzsvev^2- \snuLsvev{0}^2)     
\delta_{ik}
& \isqt \left(h^*_u \right)_{jk} \shbzsvev  
-\isqt  \mux{0}\YUs{jk} \snuLvev{0}
\\ \isqt  \left(h_u \right)_{li} \shbzvev     
- \isqt \YU{li}\muxs{0}\snuLsvev{0}
  &  \left( m_{\sUc{}{}}^2 \right)_{jm} 
+\half (Y_U Y_U^\dagger)_{jm}\shbzvev^2
-\frac{g^2}{6}(\shbzsvev^2 -\snuLsvev{0}^2) \delta_{jm}
\\
\end{array} \right) \nonumber\\
\nonumber\\[2mm]
\label{matu}
\end{eqnarray}
Recall that $(Y_U)_{ij}=\hat{Y}_{{U}i}\, \delta_{ij}$ are diagonal up
quark Yukawa couplings.

\subsection{Mass terms for down quarks}

\[\Lag \supset
-\isqt \lamp{0 ij} v_d \dL{i}{z}{} \dR{j}{z}{}
-\isqt \lamps{0 ij} v_d \dLb{i}{z}{} \dRb{j}{z}{}
\]
\begin{equation} \phantom{\Lag} =
- m_{d \, i}  \dL{i}{z}{} \dR{i}{z}{}
- m_{d \, i} \dLb{i}{z}{} \dRb{i}{z}{} \;, \qquad i=1,...3 \;.
\end{equation}

\subsection{Mass terms for up quarks}

\[\Lag \supset
- \isqt \YU{ij} \shbzvev \uR{j}{y}{} \uL{i}{y}{}
-\isqt \YUs{ij} \shbzsvev \uRb{j}{y}{} \uLb{i}{y}{}
\]
\begin{equation}\phantom{\Lag} =
-m_{u \, i} \uR{i}{y}{} \uL{i}{y}{}
-m_{u \, i}\uRb{i}{y}{} \uLb{i}{y}{} \;, \qquad     i=1,...3 \;.
\end{equation}

\subsection{Mass terms for neutrino-neutralino}

\[ {\cal L} \supset   - \half
\left( \begin{array}{ccccccc} -i \Bino{}{} & -i \Wino{}{0} & \hinobz{} &
	\nuL{\alpha}{} \\
\end{array} \right)
\M{N}{}
\left( \begin{array}{c} -i \Bino{}{} \\ -i \Wino{}{0}\\ \hinobz{} \\
	\nuL{\beta}{}\\ 
\end{array} \right) - \textup{H.c.}
\]
\[ \phantom{\Lag} = - \half
\left( \begin{array}{ccccccc} -i \Bino{}{} & -i \Wino{}{0} & \hinobz{} &
	\nuL{\alpha}{} \\
\end{array} \right)
\ZNs{} \ZNT{}
\M{N}{}
\ZN{} \ZNd{}
\left( \begin{array}{c} -i \Bino{}{} \\ -i \Wino{}{0}\\ \hinobz{} \\
	\nuL{\beta}{}\\ 
\end{array} \right) + \textup{H.c.} \]
\begin{equation}
\phantom{\Lag} = - \half \Mh{N}{pq} \Kz{p}{} \Kz{q}{}   -   \half
\Msh{N}{pq} \Kzb{p}{} \Kzb{q}{} \;, \qquad p,q=1,...7 \;.
\end{equation}
where 
\begin{eqnarray}
\M{N}{} =
\left( \begin{array}{ccccccc}
M_1    &
0    &
 \frac{g}{2}  \shbzsvev   &
 -\frac{g}{2}  \snuLsvev{0} \delta_{0 \beta} \\
0    &
M_2   &
 -\frac{g_2}{2}\shbzsvev   &
 \frac{g_2}{2}\snuLsvev{0} \delta_{0 \beta}   \\
 \frac{g}{2}  \shbzsvev   &
 -\frac{g_2}{2}\shbzsvev   &
0    &
 -\mux{\beta}   \\
 -\frac{g}{2}  \snuLsvev{0} \delta_{0 \alpha}   &
 \frac{g_2}{2}\snuLsvev{0}  \delta_{0 \alpha}  &
 -\mux{\alpha}   &
 0_{\alpha \beta}    \\
\end{array} \right) \;.  \label{matn}
\end{eqnarray}

\subsection{Mass terms for charged lepton-chargino}

\[ {\cal L} \supset   - 
\left( \begin{array}{cc} -i \Wino{}{-} & \eL{\alpha}{} \end{array} \right)
\M{C}{}
\left( \begin{array}{c} -i \Wino{}{+} \\ \hinobp{} \\ \eR{k}{} \end{array} \right)
	\hspace{10pt} - \hspace{10pt}
\left( \begin{array}{cc} i \Winob{}{-} & \eLb{\alpha}{} \end{array} \right)
\Ms{C}{}
\left( \begin{array}{c} i \Winob{}{+} \\ \hinobpb{} \\ \eRb{k}{} \end{array} \right)
\]
\[ \phantom{\Lag} = - 
\left( \begin{array}{cc} -i \Wino{}{-} & \eL{\alpha}{} \end{array} \right)
\Zm{} \Zmd{}   
	\M{C}{}
	\Zp{} \Zpd{}
	\left( \begin{array}{c} -i \Wino{}{+} \\ \hinobp{} \\ \eR{k}{} \end{array} \right)
	\hspace{10pt} 
\]
\[- \hspace{10pt}
\left( \begin{array}{cc} i \Winob{}{-} & \eLb{\alpha}{} \end{array} \right)
	\Zms{} \ZmT{}
	\Ms{C}{}
	\Zps{} \ZpT{}
	\left( \begin{array}{c} i \Winob{}{+} \\ \hinobpb{} \\ \eRb{k}{} \end{array} \right)
\]
\begin{equation}
\phantom{\Lag} = -  \Mh{C}{pq} \Km{p}{} \Kp{q}{}  -\Msh{C}{pq} \Kmb{p}{}
\Kpb{q}{} \;, \qquad p,q=1,...5
\end{equation}
where 
\begin{eqnarray}
\M{C}{} = 
\left( \begin{array}{ccc}
 M_2 &
\isqt g_2\shbzsvev &
0  \\
\isqt g_2\snuLsvev{0} \delta_{0 \alpha} &
 \mux{\alpha} &
- \isqt \lam{\alpha 0 k} \snuLvev{0} \\
\end{array} \right) \;.  \label{matc}
\end{eqnarray}

\newpage

\section{Feynman Rules for Neutral Fermions in the \LMSSM}  
\label{app:feyn}
\setcounter{equation}{0}  


\subsection{Neutral Scalar - Neutral Fermion - Neutral Fermion interactions}
\label{app:feyn1}
\noindent
\iin{\Kz{p}{}}{\Kz{r}{}}{\Hz{q}}{
-  \frac{ie}{2 c_W} \ZR{1q}  \ZN{3p}   \ZN{1r} 
+ \frac{ie}{2 s_W} \ZR{1q}   \ZN{2p}   \ZN{3r}  
}{
+  \frac{ie}{2 c_W}   \ZR{(2+\alpha)q}   \ZN{(4+\alpha)p}  \ZN{1r} 
- \frac{ie}{2 s_W} \ZR{(2+\alpha)q}  \ZN{2p}  \ZN{(4+\alpha)r} 
}{
-  \frac{ie}{2 c_W} \ZR{1q}  \ZN{3r}   \ZN{1p} 
+ \frac{ie}{2 s_W} \ZR{1q}   \ZN{2r}   \ZN{3p}  
}{
+  \frac{ie}{2 c_W}   \ZR{(2+\alpha)q}   \ZN{(4+\alpha)r}  \ZN{1p} 
- \frac{ie}{2 s_W} \ZR{(2+\alpha)q}  \ZN{2r}  \ZN{(4+\alpha)p} 
}{}\\
\iin{\Kz{p}{}}{\Kz{r}{}}{\Az{q}}{
-  \frac{e}{2 c_W}  \ZA{1q} \ZN{3p}  \ZN{1r} 
+ \frac{e}{2 s_W} \ZA{1q}  \ZN{2p}  \ZN{3r}
}{
+  \frac{e}{2 c_W} \ZA{(2+\alpha)q}  \ZN{(4+\alpha)p}  \ZN{1r} 
- \frac{e}{2 s_W} \ZA{(2+\alpha)q}  \ZN{2p} \ZN{(4+\alpha)r}  
}{
-  \frac{e}{2 c_W}  \ZA{1q} \ZN{3r}  \ZN{1p} 
+ \frac{e}{2 s_W} \ZA{1q}  \ZN{2r}  \ZN{3p}
}{
+  \frac{e}{2 c_W} \ZA{(2+\alpha)q}  \ZN{(4+\alpha)r}  \ZN{1p} 
- \frac{e}{2 s_W} \ZA{(2+\alpha)q}  \ZN{2r} \ZN{(4+\alpha)p}  
}{}\\

\subsection{Charged Scalar - Neutral Fermion - Charged Fermion interactions}
\label{app:feyn2}

\noindent
\iisi{\Km{p}{}}{\Kz{r}{}}{\Hp{q}}{
-i \frac{e}{s_W}\ZH{(2+\alpha)q}  \Zms{1p}  \ZN{(4+\alpha)r}
}{
-i \lam{\alpha \beta j} \ZH{(5+j)q}  \Zms{(2+\beta)p}  \ZN{(4+\alpha)r} 
}{
+i  \frac{e}{\sqrt{2} c_W}  \ZH{(2+\alpha)q} \Zms{(2+\alpha)p}  \ZN{1r} 
}{
+i \frac{e}{\sqrt{2} s_W}\ZH{(2+\alpha)q} \ZN{2r}  \Zms{(2+\alpha)p} }{}\\
\iiso{\Kp{p}{}}{\Kz{r}{}}{\Hps{q}}{
-i \lam{\alpha \beta j} \ZHs{(2+\beta)q}  \Zp{(2+j)p}  \ZN{(4+\alpha)r}
}{
-i  \sqrt{2} \frac{e}{c_W}  \ZHs{(5+i)q}  \Zp{(2+i)p} \ZN{1r} 
}{
-i  \frac{e}{\sqrt{2} c_W}  \ZHs{1q}  \Zp{2p}  \ZN{1r} 
-i \frac{e}{s_W}\ZHs{1q}  \Zp{1p} \ZN{3r}  
}{
-i \frac{e}{\sqrt{2} s_W} \ZHs{1q} \ZN{2r} \Zp{2p}}{}{}\\

\subsection{Squark - Neutral Fermion - Quark interactions}
\label{app:feyn3}

\noindent
\iiso{\Kz{p}{}}{\dL{j}{y}{}}{\sds{q}{y}}{
-i \lamp{\alpha jk} \Zsd{(3+k)q}  \ZN{(4+\alpha)p}    
-i \frac{e}{3\sqrt{2} c_W} \Zsd{jq}   \ZN{1p}
}{
+i \frac{e}{\sqrt{2} s_W}\Zsd{jq} \ZN{2p}     
}{}{}{}\\
\iisi{\Kz{p}{}}{\dR{j}{y}{} }{\sd{q}{y}}{
-i \lamp{\alpha ij} \Zsds{iq}  \ZN{(4+\alpha)p}   
-i \frac{e\sqrt{2}}{3 c_W} \Zsds{(3+j)q}   \ZN{1p} 
}{}{}{}{}\\
\iiso{\Kz{p}{}}{\uL{j}{y}{}}{\sus{q}{y}}{
   -i   \frac{e}{3\sqrt{2} c_W} \Zsus{jq}   \ZN{1p}
-i \YU{jk} \Zsus{(3+k)q}    \ZN{3p} 
}{
-i \frac{ e}{\sqrt{2} s_W}\Zsus{jq} \ZN{2p}  
}{}{}{}\\
\iisi{\Kz{p}{}}{\uR{k}{y}{}}{\su{q}{y}}{
i \frac{2e\sqrt{2}}{3c_W} \Zsu{(3+k)q}    \ZN{1p}
-i \YU{ik} \Zsu{iq}    \ZN{3p}  
}{}{}{}{}\\

\subsection{Fermion - Fermion - Gauge boson interactions}
\label{app:feyn4}

\noindent
\ffv{\Kzb{r}{}}{\Kz{q}{}}{\Z{\mu}}
    { \frac{i e}{2 s_W c_W} \bigg[ \ZNs{(4+\alpha)r} \ZN{(4+\alpha)q} 
    - \ZNs{3r} \ZN{3q} \bigg] \sigb{}{\mu}  }{}{}{}{} \\
\ffv{\Kzb{r}{}}{\Km{q}{}}{\W{\mu}{+}}
    { i \bigg[ \frac{e}{\sqrt{2} s_W} \ZNs{(4+\alpha)r} \Zms{(2+\alpha)q} 
    + \frac{e}{s_W} \ZNs{2r} \Zms{1q} \bigg] \sigb{}{\mu}   }{}{}{}{} \\
\ffv{\Kzb{r}{}}{\Kp{q}{}}{\W{\mu}{-}}
    { i \bigg[ \frac{e}{\sqrt{2} s_W} \ZNs{3r} \Zp{2q} 
    - \frac{e}{s_W} \ZNs{2r} \Zp{1q} \bigg] \sigb{}{\mu}   }{}{}{}{} \\


\newpage

\section{Weyl spinors and  self-energy one-loop corrections}  
\label{app:weyl}
\setcounter{equation}{0}  

Throughout this article we make an extensive use of Weyl spinor
notation.  Generally speaking, working with Weyl spinors is
advantageous because of two reasons : first they appear naturally in a
supersymmetric Lagrangian (no extra work is required to make the
connection with Dirac or Majorana four-component spinors and their
corresponding Feynman rules) and second, when used in Feynman
diagrams, they present transparently their structure as for example,
the dominance of a particle mass or the appearance of a mixing etc.
In this Appendix we give a pedagogical introduction to the use of Weyl
fermion propagators and vertices.  We then calculate at the end the
generic self energies that appear in (\ref{con1}, \ref{con2}).  This
Appendix is complementary to the work of Ref.~\cite{preSUSY}.
   
\begin{figure}[t] 
  \centering
   \includegraphics[width=3in,bb=166 511 491 721]{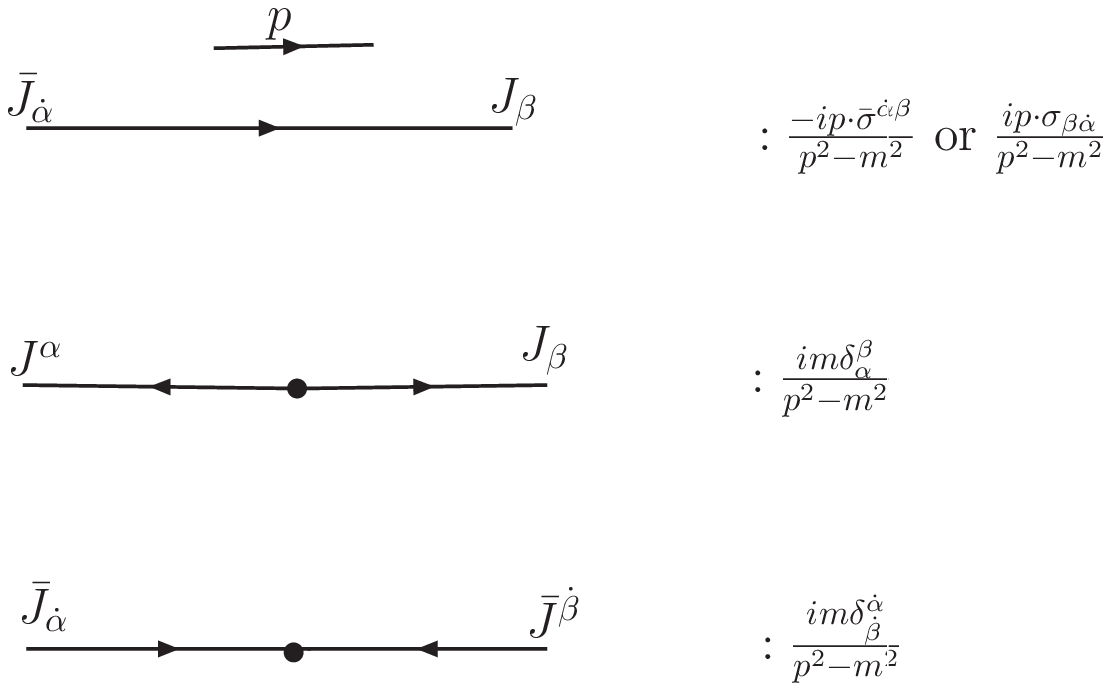} 
   \caption{Massive Weyl spinor propagators.  }
   \label{fig:prop}
\end{figure}

Using path integral technics we want to find the propagator of a
massive Weyl fermion in Minkowski space.  The path integral involving
a Weyl fermion $\xi^\alpha(x)$ and a source field $J^\alpha(x)$ is
\begin{eqnarray}
W[J, \Jb] \ = \ N \int d[\xi] \:d[\xib]\:  e^{i S[\xi, \xib, J, \Jb] }\;, \label{PI}
\end{eqnarray}
where $N$ is a constant and
\begin{eqnarray}
i \: S[\xi, \xib, J, \Jb] \ = \ i \:  \int d^4x \biggl \{ \frac{1}{2}\biggl [ i \xib \bar{\sigma}^\mu \partial_\mu \xi +
i \xi  \sigma^\mu \partial_\mu \xib - m (\xi \xi + \xib \xib) \biggr ] + J \xi +\Jb\xib \biggr \} \;.\label{AF}
\end{eqnarray}
In writing the above action functional we use the conventions of Wess
and Bagger~\cite{WB} with the metric being $g_{\mu\nu} =
(1,-1,-1,-1)$.  It is simpler to work in momentum space and thus we
Fourier transform (in four dimensions) both fields and sources as
\begin{eqnarray}
\xi^\alpha(x) \ &=& \ \int_{-\infty}^{+\infty} \frac{d^4p}{(2\pi)^2} \: e^{i p\cdot x} \: \xi^\alpha(p) \;, \qquad
\xib^{\dot{\alpha}}(x) \ = \ \int_{-\infty}^{+\infty} \frac{d^4p}{(2\pi)^2} \: e^{-i p\cdot x} \: 
\xib^{\dot{\alpha}}(p) \;, \nonumber \\[3mm]
J^\alpha(x) \ &=& \ \int_{-\infty}^{+\infty} \frac{d^4p}{(2\pi)^2} \: e^{i p\cdot x} \: J^\alpha(p) \;, \qquad
\Jb^{\dot{\alpha}}(x) \ = \ \int_{-\infty}^{+\infty} \frac{d^4p}{(2\pi)^2} \: e^{-i p\cdot x} \: 
\Jb^{\dot{\alpha}}(p) \;,
\end{eqnarray}
where $p\cdot x \equiv p_\mu x^\mu$.  We also make use of the four
dimensional definition of the $\delta$-function,
\begin{eqnarray}
\delta(x-x')  \ = \ \int_{-\infty}^{+\infty} \frac{d^4p}{(2\pi)^4} \: e^{i p\cdot (x-x')} \: .
\end{eqnarray}
The action functional (\ref{AF}) is conveniently written in a matrix
notation
\begin{eqnarray}
i \: \int d^4p \: \frac{1}{2} \biggl ( \Omega^\dagger {\cal M} \Omega 
+ \Omega^\dagger X + X^\dagger \Omega
\biggr ) \;,\label{AF2}
\end{eqnarray}
where
\begin{eqnarray}
\Omega(p) \equiv   
\left ( \begin{array}{c} \xi_\alpha(-p) \\[3mm] \bar{\xi}^{\dot{\alpha}}(p) \end{array} \right )
\;, \quad X(p) \equiv   \left ( \begin{array}{c} \Jb^{\dot{\alpha}}(p) \\ [3mm]J_\alpha(-p) \end{array} \right )
\;, \quad {\cal M}(p) \equiv \left ( \begin{array}{cc} 
\bar{\sigma} \cdot p  &  -m \\[3mm]
-m & \sigma \cdot p \end{array} \right ) \;, \label{c6}
\end{eqnarray}
where $\sigma \cdot p = \sigma^\mu_{\alpha\dot{\beta}}\: p_\mu$ and
$\bar{\sigma} \cdot p = \bar{\sigma}_\mu^{\dot{\alpha}\beta} \: p^\mu$
and the matrix ${\cal M}$ is hermitian.  The path integral measure
does not change when transforming the Weyl fermions $\xi$ by a
constant $(J)$ as
\begin{eqnarray}
\Omega \ = \ \Omega' - {\cal M}^{-1}\:  X \;.
\end{eqnarray}
After this transformation is applied in (\ref{AF2}), sources and
fields ``get decoupled''
\begin{eqnarray}
i \: \int d^4p \: \frac{1}{2} \left ( \Omega^{'\dagger} {\cal M} \Omega' - X^\dagger {\cal M}^{-1} X
\right ) \;,\label{AF3}
\end{eqnarray} 
where the inverse of {\cal M} is easily found to be
\begin{eqnarray}
{\cal M}^{-1} \ = \ \frac{1}{p^2 - m^2} \: \left ( \begin{array}{cc} \sigma\cdot p & m \\[3mm]
m & \bar{\sigma}\cdot p \end{array} \right ) \;.
\end{eqnarray} 
The first integrand term in (\ref{AF3}) is exactly the same as the one
in square brackets of (\ref{AF}).  Recalling that $X=X[J,\Jb]$ the
path integral in (\ref{PI}) takes the form
\begin{eqnarray}
W[J, \Jb] \ = \  N e^{i S_0[\xi,\xib ]} \: \exp \left \{ -\frac{i}{2} \int d^4p \: X^\dagger {\cal M}^{-1} X\right \}
= W[0]   \: \exp \left \{ -\frac{i}{2} \int d^4p \: X^\dagger {\cal M}^{-1} X\right \} \;,\nonumber \\[2mm]
\end{eqnarray}
where $S_0$ is the free Weyl fermion action functional
\begin{eqnarray}
S_0[\xi, \xib] \ = \ \int d^4x \left [
i \xib \bar{\sigma}^\mu \partial_\mu \xi  - \frac{1}{2} m (\xi \xi + \xib \xib) \right ] \;.
\end{eqnarray}
The propagators for the Weyl fermions can be read from
\begin{eqnarray}
  \exp \left \{ -\frac{i}{2} \int d^4p \: X^\dagger {\cal M}^{-1} X\right \}  &=& 
-\frac{1}{2} \int d^4p \left \{ J^\beta(p) \frac{i \sigma_{\beta \dot{\alpha}} \cdot p }{p^2-m^2} 
 \Jb^{\dot{\alpha}}(p) + \Jb_{\dot{\alpha}}(p) \left [ \frac{-i \bar{\sigma}^{\dot{\alpha}\beta} \cdot p}{p^2-m^2} \right ]J_\beta(p) \right.\nonumber \\[3mm] 
 &+& \left.  J^\alpha(p) \frac{i m \delta_\alpha^\beta}{p^2-m^2} J_\beta(-p) +
 \Jb_{\dot{\alpha}}(-p) \frac{i m \delta^{\dot{\alpha}}_{\dot{\beta}}}{p^2-m^2} \Jb^{\dot{\beta}}(p) \right \}\;.
 \label{props}\end{eqnarray}
Diagrammatically the propagator of a massive Weyl fermion is depicted
in Fig.~\ref{fig:prop}.
The convention we adopt here is that arrows run away from dotted
indices at a vertex and towards undotted indices at a vertex.  As it
is obvious from (\ref{props}), the kinetic part of the propagator [the
top one in Fig.~\ref{fig:prop}] is uniquely defined from the height
of the indices that link this propagator with the vertex.
 
\begin{figure}[t] 
   \centering
   \includegraphics[width=5in,bb=71 367 724 720]{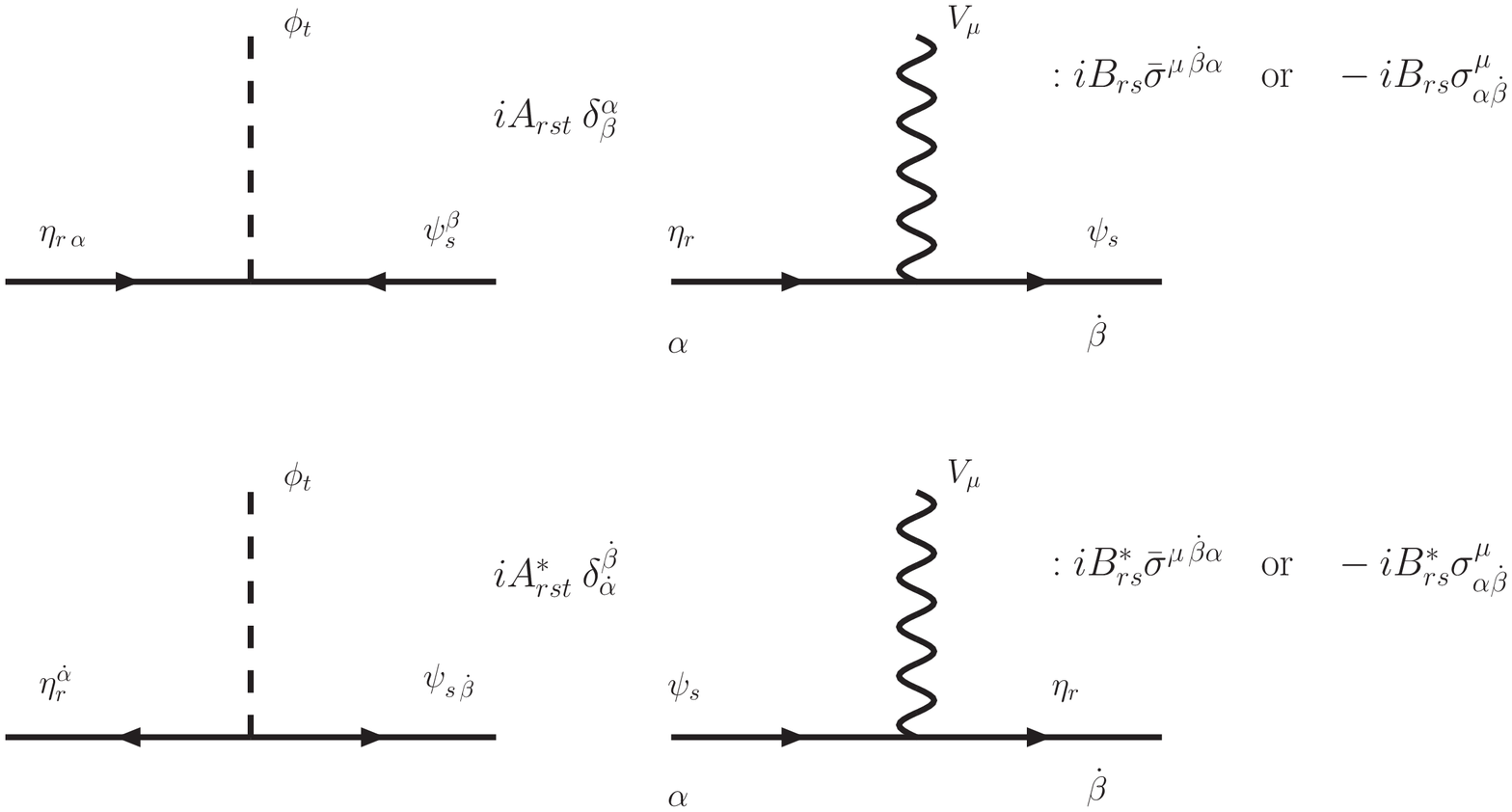} 
   \caption{General vertices involving Weyl spinors used in this
   article.  In case of the vector boson vertex the Feynman rule is
   defined completely from the height that link a propagator with this
   vertex.  Vertices on the left arise from ${\cal L}_Y = A_{rst}
   \phi_t \psi_s \eta_r + A_{rst}^* \phi_t^* \bar{\psi}_s \bar{\eta}_r$
   and vertices on the right arise from ${\cal L} = B_{rs}
   \bar{\psi}_s \bar{\sigma}^\mu V_\mu \eta_r + B^*_{rs} \bar{\eta}_r
   \bar{\sigma}^\mu V_\mu^* \psi_s$.}
   \label{fig:ver}
   \end{figure}

The propagator for two different Weyl spinors $\eta$ and $\psi$
(forming a Dirac spinor in four component notation) with action
functional,
\begin{eqnarray}
i \: S[\xi, \xib, J, \Jb]   &=&  i \:  \int d^4x \biggl \{ \biggl [
i \:  \bar{\eta} \bar{\sigma}^\mu \partial_\mu \eta 
+ i \: \bar{\psi} \bar{\sigma}^\mu \partial_\mu \psi 
- m (\eta \psi + \bar{\eta} \bar{\psi}) \biggr ] 
 + J_\eta \eta  +J_{\bar{\eta}} \bar{\eta} +J_\psi \psi  + J_{\bar{\psi}} \bar{\psi}  \biggr \} 
 \nonumber \\ \;.\label{AF4}
\end{eqnarray}
have the same form as in (\ref{props}) with obvious Lorentz invariant
substitutions.  Vertices with Weyl fermions arise from the
superpotential and from the supersymmetric gauge interactions.  They
are displayed in Fig.~\ref{fig:ver}.

We now proceed in calculating the general self energies of
(\ref{con1},\ref{con2}).  The 1PI self energy, $\Sigma^D$, obtains
corrections from diagrams which have either gauge particles or scalar
particles in the loop.  The scalar contributions for general
scalar-fermion vertices $i A_{qrs}$ and $i B_{prs}$ is
\begin{multicols}{2}
	\center
	\includegraphics[bb=109 641 242 710]{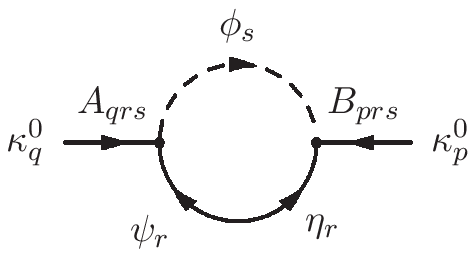}
	\columnbreak
	\flushleft
	$$
\hspace*{-1.5cm} 	\Sigma^D_{pq}(m^2_{\Kz{q}{}})=$$
$$ 
\hspace*{-1.5cm} i B_{prs} A_{qrs} m_{\psi\eta_r} \frac{i \pi^2}{(2\pi)^4 \mu^{4-D}}
	B_0(m^2_{\Kz{q}{}},m^2_{\phi_s},m^2_{\psi\eta_r}) \;,
	$$
	\begin{equation}
	\label{c1}
	\end{equation}
\end{multicols}	
\noindent where $m_{\psi\eta_r}$ denotes the physical mass of the mass
eigenstate
which is composed out of the interaction eigenstates $\psi_r$ and
$\eta_r$.  The corresponding neutrino self energy arising from vector
boson contributions with generic vertices $i C_{qr} \bar{\sigma}^\mu$
and $i D_{qr} \bar{\sigma}^\mu$ is
\begin{center}
	\includegraphics[bb=109 641 242 712]{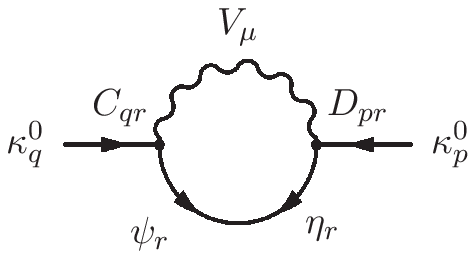}
	\begin{eqnarray}
	\Sigma^D_{pq}(m^2_{\Kz{q}{}}) &=&
	 i D_{pr} C_{qr}  m_{\psi\eta_r} \frac{i \pi^2}{(2\pi)^4 \mu^{4-D}}
	\Bigg\{ 
	(\xi+3) B_0(m^2_{\Kz{q}{}},m^2_{V},m^2_{\psi\eta_r})+ \nonumber  \\[2mm]
	&+& (\xi-1) \xi m^2_{V} C_0(0,m^2_{\Kz{q}{}},m^2_{\Kz{q}{}},m^2_{V},
	\xi m^2_{V},m^2_{\psi\eta_r}) \Bigg\} \;, \label{c2}
	\end{eqnarray}
\end{center}	
where $\xi$ is the gauge fixing parameter, $m_V$ is the mass of the
vector boson and $B_0, C_0$ are the Passarino-Veltman
functions~\cite{PV} in the notation of Ref.~\cite{Hahn},
\begin{eqnarray}
B_0(q^2, m_\phi^2, m_\psi^2) &=& \frac{(2\pi)^4 \mu^{4-D}}{i \pi^2} \int \frac{d^D k }{(2 \pi)^D} 
\frac{1}{(k^2-m_\phi^2) \: ([q+k]^2-m_\psi^2) }\;, \\[3mm]
C_0(0,q^2,q^2,m_V^2,\xi m_V^2,m_\psi^2) &=& 
 \frac{(2\pi)^4 \mu^{4-D}}{i \pi^2} \int \frac{d^D k }{(2 \pi)^D} \frac{1}{(k^2-m_V^2) \: (k^2-\xi m_V^2)
 \: ([q+k]^2-m_\psi^2)} \;.  \nonumber \\[2mm]
\end{eqnarray}
\noindent Finally, self energy corrections to the Weyl fermion kinetic
terms read as
\begin{multicols}{2}
\center
\includegraphics[bb=109 633 242 710]{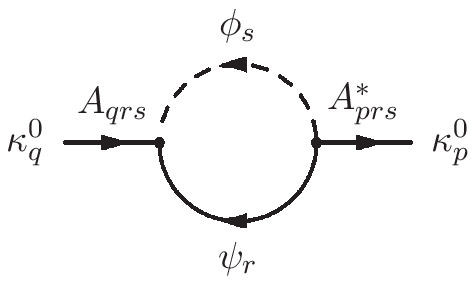}
\columnbreak
\flushleft
$$
\hspace*{-1.5cm} \Sigma^L_{pq} (m^2_{\Kz{q}{}})= i A^*_{prs} A_{qrs}
\frac{i \pi^2}{(2\pi)^4 \mu^{4-D}} B_1
(m^2_{\Kz{q}{}},m^2_{\psi_r},m^2_{\varphi_s}) \;,
$$
\begin{equation}
\label{c3}
\end{equation}
\end{multicols}
and
\begin{multicols}{2}
\center
\includegraphics[bb=109 633 242 712]{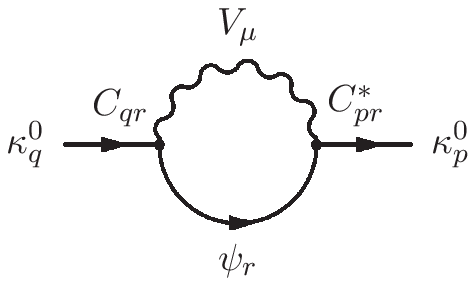}
\columnbreak
\flushleft
$$\hspace*{-1cm}
\Sigma^L_{pq} (m^2_{\Kz{q}{}}) = i C^*_{pr} C_{qr} \frac{i \pi^2}{(2\pi)^4 \mu^{4-D}}
$$
$$\hspace*{-1cm} \Bigg\{ -(\xi+1)
B_0(m^2_{\Kz{q}{}},m^2_{V},m^2_{\psi_r}) - 2
B_1(m^2_{\Kz{q}{}},m^2_{V},m^2_{\psi_r})
\Bigg.
$$
$$ \hspace*{-1cm} \Bigg.
- (\xi-1) \xi m^2_{V} C_0(0,m^2_{\Kz{q}{}},m^2_{\Kz{q}{}},m^2_{V},
\xi m^2_{V},m^2_{\psi_r})\Bigg.
$$
$$ \hspace*{-1cm} \Bigg.
- (\xi-1) (m^2_{\Kz{q}{}}-m^2_{\psi_r}) C_2(0,m^2_{\Kz{q}{}},m^2_{\Kz{q}{}},m^2_{V},
	\xi m^2_{V},m^2_{\psi_r})
\Bigg\} \;,
$$
\begin{equation}
\label{c4}
\end{equation}
\end{multicols}
where $B_1, C_2$ are defined in~\cite{Hahn}.

\end{document}